\documentclass{cplslarge}
\usepackage[authoryear]{natbib}
\usepackage{makeidx}
\usepackage{url}
\usepackage{textcomp}
\makeindex
\usepackage{amsmath} 
\usepackage{amssymb} 
\usepackage{color}

\newcommand{\ac}[1]{\textcolor{black}{#1}}

\author{S.~CHARNOZ, R.~M.~CANUP, A.~CRIDA, and L.~DONES}

\begin{document}
\renewcommand{\thechapter}{\arabic{chapter}}    
\setcounter{chapter}{17}
\chapter{The Origin of Planetary Ring Systems}
\let\svthefootnote\thefootnote
\let\thefootnote\relax\footnotetext{---\\This chapter is from the book Planetary Rings Systems, edited by Matthew S. Tiscareno and Carl D. Murray. This version is free to view and download for personal use only. Not for redistribution, re-sale or use in derivative works. \copyright Cambridge University Press, \url{www.cambridge.org/9781107113824}\,.}
\let\thefootnote\svthefootnote

\section{Introduction}

The origin of planetary rings is one of the least understood
processes related to planet formation and evolution. Whereas
rings seem ubiquitous around giant planets, their great diversity of
mass, structure and composition is a challenge for any formation
scenario. Satellite destruction by cometary impacts and
  meteoroid bombardment seem to be key processes leading to
the very low-mass rings of Uranus, Neptune and Jupiter. By contrast,
moon destruction is unlikely to produce Saturn's much more massive
rings recently, so they still represent a strong challenge for
astronomers.

Recent advances in our understanding of ring and satellite formation
and destruction suggest that these processes are closely
interconnected, so that rings and satellites may be two
aspects of the same geological system. Indeed, rings may not
be only beautiful planetary ornaments, but, possibly, an essential
step in the process of satellite formation, at least for the
  small and mid-sized moons. These recent advances have taken
advantage of the many tantalizing results from the
Cassini mission, as well as advances in numerical simulation
techniques. However, no single theory seems able to explain the origin
of the different planetary rings known in our Solar System, and it now
seems evident that rings may result from a variety of processes like
giant collisions, tidal stripping of comets or satellites, as well as
planet formation itself.  Understanding rings appears to be an
important step toward understanding the origin and evolution of
planetary environments.
 
Most work on the origin of rings has been devoted to Saturn, and
somewhat less to the rings of Jupiter, Uranus and Neptune. So our
chapter will be mainly focused on the case of Saturn. However,
processes that are common to all rings or particular to those of
Saturn will be clearly delineated. In order to build any theory of
ring formation it is important to specify physical processes
that affect the long-term evolution of rings, as well as to describe
the different observations that any ring formation model should
explain. This is the topic of section~\ref{sec:proc}. In
section~\ref{sec:Sat}, we focus our attention on Saturn's rings and
their main properties, and then discuss the pros and cons of a series
of ring formation models. We also discuss the link between rings and
satellites. In section~\ref{sec:JUN}, we extend the discussion to the
other giant planets (Jupiter, Uranus, and
Neptune). Section~\ref{sec:other} is devoted to new types of rings
--\,the recent discovery of rings orbiting small outer Solar
System bodies (Centaurs), and the
possible rings around extrasolar planets (``exo-rings''). In
section~\ref{sec:conclu}, we conclude and try to identify critical
observations and theoretical advances needed to better understand the
origin of rings and their significance in the global evolution of
planets.

\section{Ring Processes}
\label{sec:proc}

\subsection{Basics of Ring Dynamics}

Rings, as we know them in our Solar System, are disks of solid
particles, in contrast to protoplanetary disks, which have a gaseous
component. With the exception of tenuous ``dust'' rings, which can
extend far from their planets, planetary rings occur within $\approx
2.5$ planetary radii, a location prone to intense tidal forces. In
dense rings, particles have nearly-keplerian
orbits\footnote{Because a planet's rotation makes it
  somewhat nonspherical, the potential felt by a ring particle is not
  exactly that of a point mass at the center of the planet. The
  fractional difference $f$ from a keplerian potential is of order
  $J_2 (R/r)^2$, where the zonal harmonic coefficient $J_2$ is in the
  range 0.004--0.016 for the four giant planets, $R$ is the equatorial
  radius of the planet, and $r$ is the distance of the ring from the
  center of the planet. Thus $f = \mathcal{O}(10^{-3}-10^{-2}$) for planetary
  rings.},  and they cross the midplane of the rings twice per orbit, with
a vertical component of the velocity equal to the orbital velocity
times the sine of their inclination angle. The link between the
thickness $H$ of a ring and the relative velocity of particles
$\sigma_v$ is therefore $\sigma_v=H \Omega$, where $\Omega$ is the
keplerian angular velocity. By analogy between the agitation of the
particles and that of the molecules of a gas, one refers to a
dynamically \emph{cold} system when $\sigma_v$ is small compared to
the orbital velocity $V_{\rm orb}$, and to a dynamically \emph{hot}
system otherwise. With $V_{\rm orb}=r\Omega$ (where $r$ is the
distance to the center of the planet), the ratio $\sigma_v/V_{\rm
  orb}=H/r\equiv h$ is the aspect ratio.

Unlike in a gas, collisions in a debris disk dissipate energy. Hence,
if collisions are frequent (as is the case in Saturn's rings, but not
in debris disks around stars), the relative velocity drops quickly,
and the ring becomes thin and dynamically \emph{cold}. The aspect
ratio $h$ of Saturn's rings is about $10^{-7}$, making this system the
thinnest natural structure known and the best example of a
dynamically cold disk. The local physical thickness $H$
of Saturn's main rings is generally about 10~m
  \citep{zebker1984, colwellchapter2009}, comparable to the physical
size of the largest ring particles themselves
\citep{ferrari_2013}. This means that Saturn's rings are probably as
cold as possible, \emph{i.e.,} they have reached their minimum state
of internal energy. This is analogous to a thermodynamically evolved
system that has evolved over many cooling timescales.

In a dynamically hot system, the relative velocities are much larger
than the escape velocity from the surface of an object in the
system. Hence, gravitational focusing has little effect, collisions do
not permit accretion, and self-gravity has a negligible effect. By
contrast, in a dynamically cold system, particles significantly
deflect each other's trajectories, allowing for more collisions,
potentially allowing accretion. To quantify the effects of
self-gravity in a near-keplerian disk, the $Q$ parameter is used
\citep{Toomre-1964}:
\begin{equation}
Q = \frac{\Omega\,\sigma_r}{3.36\,G\,\Sigma},
\label{eq:Toomre}
\end{equation}
where $\sigma_r$ is the particles' radial velocity dispersion, $G$ is
the gravitational constant, and $\Sigma$ is the surface density of
solids. $Q$ is the ratio of dispersive forces to gravitational
forces. For $Q>2$, self-gravity can be neglected. When $Q<2$, spiral
density wakes can appear, and when $Q<1$, the system is
gravitationally unstable and clumping is expected.

However, even in a gravitationally unstable system, there can be
forces that oppose clumping of solid particles -- tides from the
central body. Tidal forces are a differential effect of gravitation
that tend to stretch any object in the gravitational field of another
one, along the axis between the two centers of mass\footnote{For
  example, it is well known that the Earth's oceans are elongated
  along an axis pointing (roughly) toward the Moon due to our
  satellite's gravity.}. If tides are very strong (because the object
exerting them is very massive, or the distance to this object is very
small), they affect the shape of the object being perturbed.  Tides
can even be as strong as the gravity at the surface of an object. In
the absence of internal strength, the object is
destroyed. \citet{roche1849} studied the deformation of a liquid blob
in orbit around a planet, and found that there is no equilibrium
solution (\emph{i.e.,} the liquid blob is dispersed), if it is closer
to the center of the central planet than
\begin{equation}
r_R = 2.45\ R_p\,(\rho_p/\rho_l)^{1/3},
\label{eq:Roche}
\end{equation} 
where $R_p$ is the radius of the planet, $\rho_p$ is its density, and
$\rho_l$ is the density of the liquid. This limit is called the Roche
radius. One can compute it in various ways (considering the separation
of two spheres, for instance, or the loss of a test particle from the
surface of a rigid object), but this only changes the leading
numerical coefficient from $\simeq 1.5$ to $\simeq 2.5$
\citep{weidenschilling1984}. The notion of the Roche limit is robust,
whereas its precise location may depend on the physical process to
consider (accretion, destruction, splitting, etc.), as well as on the
material density of the ring particles.  As the ratio of the densities
only enters Eq.~(\ref{eq:Roche}) as the one-third power, the Roche
radius is roughly 2.5 planetary radii if the ring material's density
is comparable to the planet's density.

Taking the density of porous ice for the material constituting the
rings \ac{($\approx 800$~kg/m$^{3}$)}, one finds that the Roche limit
around Saturn is at $r_R\approx 140,000$~km, near the clumpy F
ring. Hence, Saturn's rings, dominated by water ice, can never
aggregate and form a single moon: they are inside their Roche
radius. Such a Roche-interior disk is very interesting because
self-gravity can be the dominant process, but yet the ring structure
persists and no permanent accretion is possible.

\subsection{Spreading of Rings}
\label{subsec:Spreading}

Whenever two particles in a ring interact, the total angular momentum
is conserved. If the interaction changes their velocity vector, they
may exchange angular momentum. As inner particles have a larger
velocity than the outer ones, interactions generally result in a
transfer of angular momentum from the inner to the outer ones. This is
similar to a sheared viscous fluid: friction between faster and slower
rings tends to slow the former down, and accelerate the latter. Even
if rings are not fluid, and do not have a viscosity in the strict
meaning of this word, the angular momentum exchanges, being
proportional to the shear, can be modeled by a viscosity effect. In
keplerian dynamics, the viscous torque exerted by the region inside a
circle of radius $r_0$ on the outside is given by \citep[see
  e.g.][]{Pringle1981}\,:
\begin{equation}
\Gamma_\nu = 3\pi\,\Sigma\,\nu\, {r_0}^2\,\Omega_0\ ,
\label{eq:Tnu}
\end{equation}
where $\nu$ is the kinematic viscosity, and $\Omega_0$ is the orbital
frequency at $r_0$.

Angular momentum conservation and mass conservation combined yield the
variations of the surface density $\Sigma$ \citep[see,
  e.g.,][]{Salmon-etal-2010,Pringle1981}:
\begin{equation}
\frac{\partial \Sigma}{\partial t} = \frac3r\,\frac{\partial}{\partial
  r}\left[\sqrt{r}\frac{\partial (\nu\Sigma\sqrt{r})}{\partial
    r}\right]
\label{eq:dSdt}
\end{equation}

Using a constant, uniform viscosity $\nu$, \citet{LBP74} derived the
equations for the evolution of an initially infinitesimally narrow
gaseous disk. However, the equation driving any astrophysical disk
controlled by viscosity and gravity is formally the same, so it is
also used in the context of planetary ring evolution. As the innermost
regions orbit faster, they transfer angular momentum to the outer
ones, which increases the orbital radius of the outer material, and
decreases that of the inner material.  The ring spreads, while angular
momentum flows outwards. The frontier separating inward accretion and
outward spreading moves outwards with time. The theoretical final
state is that of lowest energy\,: all the mass has fallen to the center,
while all the angular momentum is carried by an infinitesimally small
particle at infinity.

In Roche-interior rings, the interaction between particles cannot be
modeled by a constant, uniform viscosity.  Ring particles exchange
angular momentum when they collide, but also due to their
gravitational interaction during close encounters. The larger the
surface density of the rings (i.e., the smaller the Toomre parameter
$Q$), the stronger these exchanges can be. \citet{Daisaka-etal-2001}
provide a complete description of this phenomenon and a complex
prescription for the viscosity $\nu$, which depends on $Q$ and
$\Sigma$.

\begin{figure}
\figurebox{}{10pc}{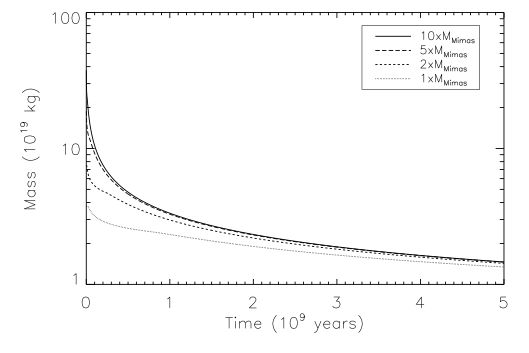}
\caption{Mass of Saturn's rings as a function of time, starting with
  different initial values of the rings' mass. The material flows
  either into Saturn's atmosphere or leaves the Roche limit. Whatever
  the initial mass, the final mass is always nearly the same,
  after 2 Gyr of evolution. The mass is controlled by the
  value of $Q\simeq 2$ everywhere. This calculation includes
    the ring mass lost to the satellite system. Adapted from
  \citet{Salmon-etal-2010}.}
\label{fig:ringmass}
\end{figure}

\begin{figure}
\figurebox{}{10pc}{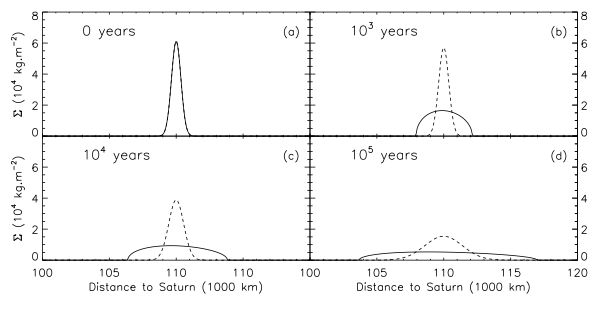}
\caption{Ring surface density at different evolution times for
  variable (solid line) and constant (dashed line) viscosities. (a)
  Initial profile. (b) At $1000$ years of evolution. (c) At $10^4$
  years of evolution. (d) At $10^5$ years of evolution. The disk with
  variable viscosity spreads faster and does not keep the original
  shape of the density profile. Adapted from
  \citet{Salmon-etal-2010}.}
\label{fig:ringevol}
\end{figure}

According to \citet{Daisaka-etal-2001}, in the gravity-dominated
regime (massive rings), the viscosity $\nu$ of the rings is
proportional to their mass squared. The characteristic time for
viscous spreading of a ring of radius $r_R$ is $t_\nu = {r_R}^{2}/\nu
\approx \frac{1}{30}\mu^{-2}T_R$, where $\mu=M_{\rm rings}/M_p$ (where
$M_{\rm rings}\approx\pi\Sigma\,{r_R}^2$ is the mass of the rings and
$M_p$ is the mass of the planet), and $T_R$ is the orbital period at
$r_R$. As $d\mu/dt = -\mu/t_\nu$, one finds $d\mu/d\bar{t} =
-30\,\mu^3$, with $\bar{t}=t/T_R$. The solution of this differential
equation is \ac{\citep{Crida-Charnoz-2014}}\,:
\begin{equation}
\mu(t)=\frac{1}{\sqrt{60\,\bar{t}+\mu_0^{\,-2}}}\ .
\label{eq:D(t)}
\end{equation}
Clearly, memory of the initial ring mass through $\mu_0$ is erased
once $\bar{t}\gg 1\,/\,60\mu_0^{\,2}$\,, and in the long term,
$\mu(t)\approx (60\,\bar{t})^{-1/2}$, independent of $\mu_0$. Around
Saturn, with $t = 4.5$ billion years, this gives $\mu=8\times
10^{-8}$. Note that this is a slowly varying function, so the result
varies by only a factor of 2 from 3.5 billion years ago to the
present. \ac{This is illustrated by 1D numerical simulations of the
  evolution of the rings (Eq.~(\ref{eq:dSdt})\,) performed by
  \citet{Salmon-etal-2010} who have studied the global, long-term
  evolution of rings and of their density profile using the
  prescription of \citet{Daisaka-etal-2001} for the viscosity
  (figure~\ref{fig:ringevol}). Figure~\ref{fig:ringmass} illustrates
  the erasure of the initial conditions mentioned in the previous
  calculation\,: whatever the initial mass, the final state has a mass
  of few $\times 10^{22}$~g after 4 Gyr.}

\ac{Today, estimates of the ring mass based on surface densities
  derived from density waves (mostly in the A ring) give a mass of
  $\simeq 4\times 10^{22}$~g, corresponding to $\mu=7\times
  10^{-8}$. The agreement between the model} of ring spreading and the
measured mass suggests that Saturn's rings could be primordial, and
that their present mass may not be the result of their formation
process, but rather of their evolution during Solar System
history. \ac{In particular, it means that Saturn's rings could have
  been much more massive in the past, by almost arbitrary amounts, and
  that the initial mass of Saturn's rings has a lower limit but not an
  upper limit.} Alternatively one could argue that Saturn's rings are
much younger than the Solar System, and that the observed coincidence
between the mass of today's rings and the asymptotic mass of a
self-gravitating disk is just a matter of luck. Future measurements of
the rings' current mass, expected during the final orbits of Cassini,
as well as analysis of the current flux of meteoroid bombardment by
the Cosmic Dust Analyzer team, may provide important constraints
relevant to these issues. Nonetheless, the viscous calculations
\ac{above} allow for the possibility of an almost arbitrarily large
initial ring mass, and thus open new doors for explaining their
formation\ac{, as we will see in section~\ref{subsec:MIR}.}

Variable viscosity also helps to maintain sharp ring structures. The
decrease of $\nu$ when $\Sigma$ drops makes the ring have sharp edges,
which move slower than in the case of a uniform viscosity. The densest
regions spread faster, but the spreading slows down dramatically as
soon as the self-gravity effects, responsible for angular momentum
transfer, decrease. In fact, in the simulations of
\citet{Salmon-etal-2010}, the surface density profile of the rings
converges naturally to a profile which makes $Q=2$ everywhere (where
$Q$ is defined by Eq.~(\ref{eq:Toomre})\,). This occurs because in the
\citet{Daisaka-etal-2001} prescription, the gravitational component of
viscosity is switched off where $Q>2$, causing a drop in $\nu$ which
almost freezes the rings' profile. This behavior is typical of
self-gravitating disks where there is a feedback mechanism between
self-gravity and heating: if $Q<2$ then the disk heats up because of
the appearance of shock waves and spiral arms, thus increasing
$Q$. Conversely when $Q>2$ internal dissipation (here in collisions)
helps the disk to cool down and thus $Q$ decreases. \ac{The existence
  of an asymptotic state for Saturn's rings by maintaining $Q \simeq
  2$ everywhere is not limited to planetary rings, but seems to be a
  general property of self-gravitating disks, and is also expected for
  circumstellar disks \citep[see, e.g.,][]{Rice_Armitage_2009}.}

\begin{figure}
\figurebox{}{10pc}{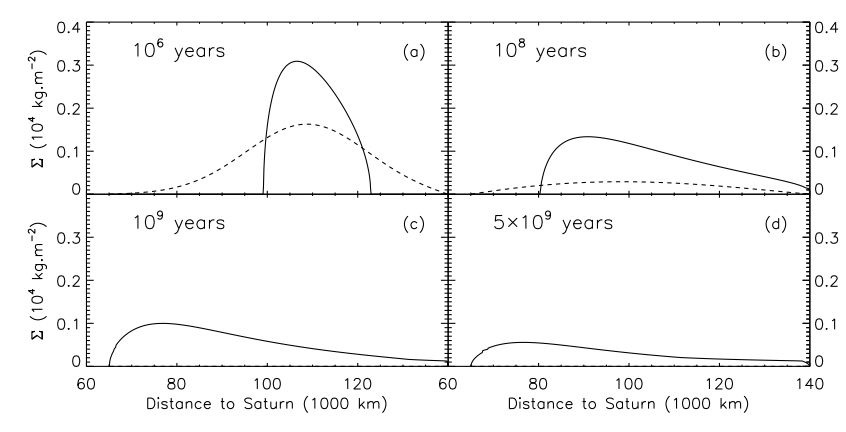}
\caption{Ring surface density at different evolution times with
  variable (solid line) and constant (dashed line) viscosities. (a) At
  1 Myr of evolution. (b) At 100 Myr of evolution. (c) At 1 Gyr of
  evolution. (d) At 5 Gyr of evolution. The disk with constant
  viscosity disperses in 1 Gyr, while the disk with variable viscosity
  remains massive over 5 Gyr with an inner density peak and 
  lower densities further out. Adapted from \citet{Salmon-etal-2010}.}
\label{fig:ringevol2}
\end{figure}

Such a profile is in qualitative agreement with the observed one, and
Saturn's rings are indeed marginally gravitationally unstable. The
``peak" seen in figure~\ref{fig:ringevol2} at $5\times 10^9$ yr could be
associated with the B ring (whereas its surface density is still a
matter of debate), and the long tail may be associated with the A
ring. Interestingly, this study suggests that there is at most a
factor 2 to 3 difference between the surface densities of the two
rings.

Of course, the detailed structure of Saturn's rings is not explained
by this simple viscous model. In particular, it does not explain the
existence of the C ring, the origin of the Cassini Division, and
numerous small-scale structures in the rings such as plateaus and
ramps. Other processes clearly sculpt the rings, including bombardment
and ballistic transport (see Chapter~\ref{Estrada}), as well
as ring-satellite interactions.

\subsection{Perturbation by Satellites}
\label{subsec:Perturb}

Exchange of angular momentum via gravity are not limited to particle
-- particle interactions. Satellites orbiting beyond the rings have a
gravitational influence on the particles, deflect their trajectories,
and on average take angular momentum from them. Computing this
deflection and integrating over all the rings,
\citet{Lin-Papaloizou-1979} found that the total torque felt by a
satellite whose mass ratio to the central planet is $\mu_s$ and
orbital radius is $r$ is given by:
\begin{equation}
\Gamma_L = \frac{8}{27}\mu_s^2\Sigma r^4\Omega^2 \Delta^{-3}
\label{eq:TL}
\end{equation}
where $\Delta = (r-r_{\rm rings})/r$ is the normalized distance
between the satellite and the outer edge of the rings, and is assumed
to be small (also see \citet{goldreich_tremaine1980}). This expression
is only valid when the satellite is close to the ring system
\citep[$\Delta \lesssim 0.1$, see][their
  Fig.~8]{MeyerVernet-Sicardy-1987}.

This formula is fundamental in ring-satellite interactions. First,
satellites close to the rings receive a positive torque, which makes
them migrate outwards, at a decreasing speed as $\Delta$ increases. As
this happens, angular momentum is removed from the rings, slowing down
their outward spreading. If the viscous torque $\Gamma_\nu$ (see
above) is smaller than this so-called Lindblad torque $\Gamma_L$,
outward spreading is thwarted and the ring's outer edge can be
confined. This is presently the case around Saturn: e.g., the outer
edge of the A ring at 136,700~km is confined by the interaction at a
7:6 resonance with the coorbital moons Janus and Epimetheus, and the
outer edge of the B ring is confined by the 2:1 resonance with
Mimas. This illustrates that if, in the history of the system, some
massive satellites have existed near the rings, they will have
perturbed the spreading of the rings and thus their overall evolution.

\subsection{Modified Accretion at the Roche Limit}
\label{subsec:MARL}

Because they are dynamically very cold, and their spreading almost
stops once $Q>2$, Saturn's rings are maintained in a marginally
gravitationally unstable state \citep[e.g.,][]{Ward1984}.
As rings spread beyond the Roche radius, they have a natural tendency
to clump, and planetary tides become too weak to prevent this natural
effect of self-gravity in a dynamically very cold system. It is
therefore expected that elongated aggregates of ring particles will
form, whose shape is the Roche lobe. The Roche lobe is the region
within which the \ac{gravitational attraction of a body dominates over
  the differential attraction} of the central planet, and its radial
width is given by the Hill radius, $R_h=a (\mu_s/3)^{1/3}$, where $a$
is the distance to the planet. Loosely bound aggregates are therefore
very fragile. A high-speed collision can deliver enough energy to
unbind the particles, but slow collisions are constructive, and as
such a rubble pile can dissipate some impact energy in deformation and
compaction. The further they migrate outwards, the larger, the more
spherical, and the more resistant to disruption such rubble piles
become. Tides can also modify two-body accretion. A particle near the
Roche limit nearly fills its Hill sphere, as the definitions of $R_h$
and $r_R$ combine into $R/R_h \approx 0.6(r_R/a)$, where $R$ is the
particle radius. Two colliding particles must remain inside their
mutual Hill sphere in order to remain
gravitationally bound to each other. This imposes severe constraints
on the geometry of encounters necessary for accretion (and the closer
to Saturn, the more severe they are). \citet{Canup_Esposito_1995} find
that tidally modified accretion between objects differing in size may
occur inside the Roche limit, but accretion between equal-sized bodies
is prevented. This results in a bimodal size distribution with big
bodies remaining on near-circular orbits (but prevented from
accreting), co-existing with clouds of tiny particles on more
eccentric and inclined orbits. Collisions between big bodies may
result in the formation of rings like Saturn's F ring close to the
Roche limit, as was shown recently \citep{Hyodo_2015}.

\section{Rings of Saturn}
\label{sec:Sat}

\subsection{Surface Density and Mass}

The mass of Saturn's rings is a key quantity that constrains models of
their origin. We do not yet have a dynamical measurement of the mass
of the entire ring system, although the Cassini radio science
experiment is expected to determine this quantity during the last year
of the mission in 2017. At present, we have measurements of the
surface mass density, $\Sigma$ of the rings at a variety of locations,
primarily in the outermost main ring, the A ring, where numerous waves
launched by nearby satellites are present. Surface densities in
regions where dynamical values are not available are sometimes
estimated by assuming that the surface density of the rings is
proportional to the optical depth, whose radial profile is measured by
occultations across the whole ring system. However, the assumed
proportionality requires that the internal density of ring particles
and their size-frequency distribution is the same in different
regions. As we discuss below, this condition is not always satisfied
in the rings \citep{tiscareno2013_aring,hedman2016}.

Saturn's satellites, particularly the ``ring moons'' Prometheus,
Pandora, Janus, Epimetheus, Atlas, Pan and Daphnis as well as the
innermost ``classical'' satellite, Mimas, perturb the rings at
locations where a resonance condition is satisfied. In most cases, the
perturbations are in the ring plane and cause tightly wound spiral
density waves. At a particularly strong resonance, a gap can form. For
instance, the Mimas 2:1 Inner Lindblad Resonance (ILR) marks the outer
edge of Saturn's B ring. The rings also harbor a few vertical
corrugations known as spiral bending waves, which are excited by moons
with (slightly) inclined orbits. As an example, both the Mimas 5:3
density wave and the Mimas 5:3 bending wave are prominent features in
the outer A ring. In this part of the A ring, ring particles complete
roughly five orbits around Saturn for every three orbits of
Mimas. However, the resonance condition also involves the precession
rate of the ring particles (positive, i.e., prograde, for apses and
negative, i.e., retrograde, for nodes). Precession splits the
resonance, so the 5:3 Inner Vertical Resonance, at which the bending
wave is excited, lies some 400~km interior to the 5:3 ILR, at which
the density wave is excited.

Density and bending waves can be used to determine the rings' surface
density $\Sigma$ because their wavelengths are proportional to
$\Sigma$. As described above, the A ring is the region where the
rings' surface density is best known, as it contains numerous
resonances with small moons. From the wave structure the derived
surface density is about 40~g/cm$^2$ \citep{Tiscareno_2007} for the A
ring. A recent study \citep{hedman2016} suggests that the B ring's
surface density $\Sigma$ is between 40 and 140~g/cm$^2$. As the
optical depth $\tau$ in much of the B ring is vastly larger than that
in the A ring (by roughly a factor of 10, see
    \citet{colwellchapter2009}), the inferred surface density of the
B ring implies a smaller value of $\Sigma/\tau$ there. This, in turn,
implies that the effective ring particle size is smaller in the B
ring, and/or the particles there have smaller internal densities than
in the A ring. \citet{hedman2016} infer that the mass of the B ring is
about $\frac{1}{2}$ to $\frac{2}{3}$ that of Mimas, and that the total
mass of the rings is, at most, comparable to that of Mimas. On the
other hand, Larry Esposito (personal communication, 2016) maintains
that the five waves studied by Hedman and Nicholson do not sample the
B ring adequately, and that the total mass of the ring system might be
significantly greater than that of Mimas
\citep{Robbins_et_al_2010}. See Chapter~\ref{Saturn} for a discussion on the rings' opacity and surface density.

\subsection{Composition and Age} 
\label{subsec:compo}

Since spacecraft have not directly sampled the particles in Saturn's
main rings, we must use reflection spectra and color to get some
indication of their composition. In general, ring particles are
similar to the nearby moons, at least on their
  surfaces. However, some small spectroscopic differences have been
  identified, and the rings are somewhat redder than the moons
  \citep{Filacchione_2014}. Saturn's rings are predominantly water
ice and therefore bright \citep{Nicholson_2008}; by comparison, the
macroscopic particles in the jovian, uranian, and neptunian rings are
dark (their albedos are small). Color variations across Saturn's rings
may indicate varying composition, possibly due in part to the effects
of the interplanetary dust that bombards them and darkens the
particles. It is likely that Saturn's ring particles have rough,
irregular surfaces resembling frost more than solid ice. There is good
indication that the particles are under-dense (internal densities $\ll
1$~g/cm$^3$, \citet{zhang_2016}), supporting the idea of ring
particles as temporary rubble piles. These slowly spinning particles
collide gently with collision velocities of just mm/sec. The
composition of Saturn's rings, with more than 90 to 95\% water ice,
seems to be in strong contrast with some of Saturn's satellites
(Titan, Dione, and Enceladus), which are approximately
half-rock, half-ice mixtures, roughly as expected for a solar
abundance of solid material. Because bombardment of the rings by
silicate-rich micrometeoroids increases their rock content over time,
the rings' current composition implies that they were essentially pure
ice when they formed.

Ring-satellite interactions can lead to evolution on timescales much
shorter than the age of the Solar System. The expected rate of
transfer of angular momentum from the rings to the satellites
(section~\ref{subsec:Perturb}) implies that several of the moons such
as Prometheus and Pandora were spawned from the rings $\ll 100$~Myr
ago \citep{goldreich1982,lissauer1984,poulet2001,Charnoz_2010}.  The
angular momentum gained by the moons is lost by the rings, and is
predicted to cause collapse of the A ring to its inner edge in
100~Myr.

The vast expanse of Saturn's rings presents an enormous target for
impact by interplanetary dust grains. These hypervelocity collisions
can erode ring particles and cause loss of ring material to Saturn's
atmosphere
\citep{cook_franklin_1970,morfill1983,northrop1987,cuzzi1990}. The
same impacts can cause the rings' structure and composition to evolve
due to the exchange of ejecta (``ballistic transport'') between
different parts of the rings \citep{ip1983,ip1984,lissauer1984,
  durisen1984, durisen1989, durisen1992, durisen1996, doyle1989,
  cuzzi1990, durisen1995, Cuzzi_Estrada_1998, latter2014, estrada2015}
[also see section 17.3.1 in the review by
  \cite{charnoz_chapter_2009}].  Even in the absence of ballistic
transport, accretion of interplanetary dust, which has a low albedo
\citep{ishiguro2013}, should darken the rings
\citep{doyle1989,elliott2011}.  Unfortunately, the timescales on which
these effects operate are uncertain, largely because the rate at which
interplanetary projectiles strike the rings \citep{tiscareno2013} is
still uncertain. See Chapter~\ref{Estrada} for a
  discussion of this issue.  Recent measurements from the Cassini
  spacecraft (the Cosmic Dust Analyzer experiment, CDA hereafter) seem
  to show that meteoritic bombardment could be intense enough so that
  it may have loaded the rings with several times their own mass over
  the age of the Solar System (see Chapter~\ref{Estrada} for more
  details). However, the CDA results remain unpublished, and the time
  variation of the bombardment rate over the lifetime of the rings is
  unknown.  If the CDA results are confirmed and their derived
  bombardment rates are viewed as typical of past rates,
  they may imply a ring age much smaller than the age of the Solar
  System. However, these results are still uncertain.

In principle, limits on the age of Saturn's rings can be derived from
observations of craters on the moons. If young ($\ll 4$~Gyr) surface
ages were derived for the moons, it would hint at young rings as well
because of the strong interactions of the rings with the satellites
out to Mimas.

The time during which craters have accumulated on a satellite can be
calculated if the rate of impact of bodies of different sizes is
known, and if the characteristics of the impactors (size, density,
velocity, etc.) can be related to the sizes of the observed craters
(see section~\ref{subsec:UN}). In recent years, Saturnian satellite
ages have been estimated from crater densities by, among others,
\citet{Zahnle_2003,kirchoff2009,dones2009}, and
\citet{disisto2016}. All of these studies assume that ecliptic comets
(also called Centaurs in the region of the giant planets) are the
primary impactors. The estimated satellite ages are generally billions
of years, with the notable exception of Enceladus's active south polar
region \citep{porco2006}. At first glance, these ``crater ages'' argue
for old rings. However, the method is not well-calibrated, because we
do not have independent knowledge of the size-frequency distribution
of the small ecliptic comets ($\mathcal{O}$(0.1--10~km diameter)\,)
that are believed to produce most of the craters seen. In fact, on the
surface of Rhea and on Dione's cratered plains, there are more
observed craters larger than 6~km than expected in 4.5~Gyr
\citep{disisto2016}\,; either the assumed population of Centaurs
should be revised (perhaps many more Centaurs struck the rings in the
early Solar System, see section~\ref{sec:CDSM}), or another reservoir
of impactors is needed.

\citet{cuk2016} investigated the dynamical evolution of the mid-sized
saturnian moons due to tides. They infer that the moons have migrated
little. Tethys and Dione probably did not cross their 3:2 resonance,
but the system likely did cross a Dione-Rhea 5:3 resonance and a
Tethys-Dione secular resonance. These crossings would have happened
recently\,: within the past 100~Myr for $Q_p = 1500$ \citep[][see
  Eq.~(\ref{eq:tides}) below]{Lainey_et_al_2012}. \citet{cuk2016}
suggested that a previous generation of moons underwent an orbital
instability, perhaps due to a solar evection resonance, leading to
collisions between them. Today's moons would have reaccreted from the
debris (at locations that allow for the Dione-Rhea 5:3 resonance
crossing, but not the Tethys-Dione 3:2), and at least the current
rings would presumably be young, although \citet{cuk2016} do not
address this issue in detail. This model implies that most craters on
the moons were formed by the debris, with impacts taking place at much
lower speeds than applies for impacts by comets. While still
speculative, this model is interesting because it opens a possibility
of forming rings recently, since collisions of icy moons might happen
at high enough velocities to be completely destructive (i.e.,
collision velocities substantially greater than the escape velocities
of the colliding bodies, \citet{movshovitz2016}). However the fate of
the debris and whether it could yield the current mid-sized satellites
and the rings (and their unique compositions) has not been
quantitatively investigated, and would appear dynamically challenging
(see section~\ref{subsec:cuk} for more details). Further, this model
requires a pre-existing system of inner mid-sized satellites, and thus
it seems plausible that there could have been a pre-existing ring
system as well.  Even if a recent large-scale dynamical modification
of the inner Saturnian system did occur, the original system mass and
compositional trends could possibly reflect a much earlier, perhaps
even primordial, epoch of formation.  Such issues merit further
consideration.

Collisions between similar-sized moons also occur in the
\citet{Crida_Charnoz_2012} model, which suggests that the regular
satellites within Titan's orbit formed from a series of giant impacts
involving roughly equal-sized bodies on nearly-circular orbits. In
  this scenario, though, the collisions are not destructive but
  constructive. Still, the debris of these collisions would generate
secondary impacts on the target, and possibly on other satellites.
Much work remains to be done to determine whether these scenarios can
be distinguished from the one discussed above, in which the largest
craters are made by comets and planetocentric debris makes only
smaller craters \citep{alvarellos2005}.

\subsection{Models for the Origin of Saturn's Rings}
\label{subsec:origSat}

That debris orbiting interior to the Roche limit at Saturn would
remain dispersed in a ring, rather than accumulating into one or more
moons, can be understood from the basic principles described above.
However, two questions have proved harder to answer: first, what was
the source of the material comprising the current main rings at
Saturn, and second, why is Saturn's ring system so much more massive
than those of the other gas giants?

During the final stages of Saturn's formation, the planet was likely
surrounded by a circumsaturnian disk containing both gas and
  solids (rock and ice).  Such a disk is a natural birthplace for
Titan, as well as some, or maybe all, of the other regular Saturnian
satellites out to Iapetus. Initially Saturn would have been much
larger than it is at present due to the energy of its accretion
\citep[e.g.,][]{pollack_1977}. But so long as the planet contracted to
within the Roche limit for ice while its disk was still present
\citep[which appears plausible,
  e.g.][]{pollack_1977,Marley_et_al_2007,Fortney_et_al_2007}, one
would expect there to have been an evolving ring of material near the
disk's inner edge.

The lifetime of a ring particle orbiting within a gaseous disk is
generally short, because the velocity differential between the
pressure-supported gas and the keplerian motion of the particle acts
as a drag on the particle's motion that causes it to spiral inward and
eventually collapse onto the planet.  Thus it seems likely that
hypothetical primordial ring systems formed within Saturn's
circumplanetary disk were lost.  Yet Saturn's rings must have formed
in an environment that allowed them to survive until the current time.
The original ring system must have had a mass comparable to or greater
than the current ring mass, $\ge$ few $\times 10^{22}$ g, and a
composition of essentially pure ice, given that the current rings are
$> 90$\% water ice, despite continuous exposure to external
bombardment that over time increases their rock content
\citep[e.g.,][also see
  section~\ref{subsec:compo}]{Cuzzi_et_al_2010,Cuzzi_Estrada_1998}.
Finally, whatever event(s) produced Saturn's rings did not produce a
comparably large and long-lived ring system around Jupiter, Uranus, or
Neptune.  It has been proposed \citep{Crida_Charnoz_2012} that Uranus
and Neptune did have early massive ring systems that were somehow lost
while Saturn's massive rings were not, as we describe in
section~\ref{subsec:UN}.  Simultaneously satisfying this set of
constraints for Saturn's rings is challenging, and a variety of origin
models have been proposed.  Below we briefly describe these models,
focusing in particular on developments occurring since the
\citet{charnoz_chapter_2009} review was published.

\subsubsection{Condensation Within a Satellite-Forming Disk}

The idea that Saturn's rings could represent material left over from
the protosatellite disk that never coagulated into a satellite, also
known as the condensation model, was developed by
\citet{Pollack_1975} and \citet{pollack_1977} \citep[see Sec. 4.1 in][for
  additional discussion]{charnoz_chapter_2009}.  In this model, the
rings are unaccreted remnants from the same disk of gas and solids
that gave rise to the regular satellites.  A first challenge is the
survival of such a ring against gas drag.  For a dense ring, drag by a
gas disk can be described as a shear stress on the disk surfaces,
resulting in a ring decay timescale
\citep[e.g.,][]{Goldreich_Ward_1973,Harris_1984}:
\begin{align}
\tau_{gd} &= 14\,Re\left ( \frac{\Sigma}{\Sigma_{g}}\right ) \left
(\frac{GM_p}{c^3}\right ) \cr &\sim 10^{2}\, \rm{yr} \left ( \frac {Re}
    {10^2} \right ) \left ( \frac{\Sigma/\Sigma_{g}}{0.03}\right )
    \left ( \frac{200 \! {\rm \; K}} {\it T}\right )^{3/2},
\label{eq:gd}
\end{align}
where $Re$ is the Reynolds number \citep[uncertain, but likely in the
  range of 50 to 500; e.g.,][]{Weidenschilling_Cuzzi_1993}, and
$\Sigma_{g}$ is the gas surface density, $c$ is the thermal velocity
of the gas molecules, and $T$ is the gas temperature.  For a disk
whose gas surface density is substantially higher than its surface
density of solids (as would be the case for a solar composition disk),
the implied lifetime of a condensed ring is then short compared to the
likely lifetime of the gas disk, which could be $\ge 10^6$ yr.

Low-temperature condensation would generally be expected to produce a
rock-ice ring reflecting bulk solar abundances, inconsistent with the
essentially pure ice composition needed for consistency with Saturn's
rings. A clever solution to this problem was proposed by
\citet{Pollack_1976}, who argued that as the disk cooled, silicates in
the inner disk would condense before the ices did.  If these silicates
were lost to gas drag decay, subsequent condensation of ices as the
disk cooled further would yield an ice ring.  A condensation origin of
Saturn's rings thus requires that (1) ice condensation is delayed
until after the earlier-formed rocky ring has been removed, and (2)
ice condensation occurred concurrently with the dispersal of the
gas disk, so that the resulting ice ring survived.  The first is
plausible given that timescales in Eqs.~(\ref{eq:gd}) may be short
compared to disk cooling timescales
\citep[e.g.][]{charnoz_chapter_2009}. However, (2) appears to require
a coincidence.

The condensation model remains an interesting idea, but the
  lack of a quantitative model of ring origin by this process makes it
  hard to assess how restrictive it would be compared to other
  contemporary ring origin ideas.  The viability of a condensation
  model would appear to depend on the nature of satellite accretion,
  the assumed lifetime and thermal evolution of the protosatellite
  disk, and considerations involving the timing and rate of gas inflow
  to the disk (also see discussion in section~\ref{sec:TSLS} below).
  Although these issues are (and will remain) uncertain, progress
  could be made by using traditional assumptions as a starting point
  to evaluable the feasibility of ring origin via condensation
  compared to other ideas that also, by necessity, rely on such
  assumptions.

\subsubsection{Tidal Disruption of a Small Moon}

In 1847, Edouard Roche suggested that Saturn's rings originated when a
small moon strayed within the Roche limit and was torn apart by tidal
forces \citep{roche1849}.  While appealing in its simplicity, this
idea has generally been disfavored of late due to the perceived
difficulty in effectively disrupting a Mimas-sized object.  Tidal
stresses on a density $\rho$, radius $R$ satellite with semi-major
axis $a$ and orbital frequency $\Omega$ are of order $T
\sim\rho\Omega^2 R^2$ \citep[e.g.,][]{weidenschilling1984}.  For small
objects, material strength inhibits tidal disruption, while for larger
objects, self-gravity dominates and strength is less important.  The
transition between these two regimes occurs at $R \sim 200$~km for
solid ice \citep{Sridhar_Tremaine_1992}, which is approximately the
size of a progenitor moon whose mass is comparable to that in the
current rings.
		
In the limit of no strength and a fluid-like, self-gravitating body,
the relevant tidal disruption distance would be the classical Roche
limit (Eq.~\ref{eq:Roche}).  However, a small, Mimas-sized ring
progenitor would likely need to orbit well within the Roche limit to
disrupt, probably interior to the inner edge of the current main rings
(\cite{jeffreys1947}; \cite{aggarwal1974}; \cite{davidsson1999}; also
see section 4.2.2 in \cite{charnoz_chapter_2009}), and perhaps even
interior to the expected early position of Saturn's surface (see
figure~\ref{fig:async} below).  If the ring progenitor was a regular
moon, it would need to evolve from an orbit substantially outside its
Roche limit (where it could have initially formed) to one well within
the Roche limit.  If this were accomplished through interactions with
a gas disk, the resulting dispersed fragments would be vulnerable to
loss via gas drag, because gas densities high enough to drive a small
moon's inward migration through gas drag would cause a much faster
destruction of its mass when spread amongst small fragments.  Finally,
complete disruption of a nominal composition satellite that contained
rock and ice in roughly solar proportions would lead to a rock-ice
ring, rather than a nearly pure ice ring.  Thus a tidal disruption
model needs to be considered in conjunction with a model for producing
an extremely rock-poor icy progenitor moon.
		
\subsubsection{Collisional Disruption of a Small Moon}
\label{sec:CDSM}
Another ring origin theory proposes that during the satellite
formation era, a $\sim 200$ km radius, few $\times 10^{22}$~g
satellite drifted inward to an orbit interior to its Roche limit,
located at $r_R = 2.45R_{S}(\rho_p/\rho)^{1/3} \approx 2.17(\rho/1\;
{\rm g \; cm^{-3}})^{-1/3} R_{S}$. In this classic
  expression, $R_{S}$ is Saturn's mean radius rather than its
  equatorial radius, with $R_{S} = 58,232$ km currently. The satellite
remains interior to the Roche limit until a later time when it is
disrupted by a heliocentric impactor
\citep[e.g.,][]{Harris_1984,charnoz_lhb_2009,charnoz_chapter_2009}.
The collisional disruption model addresses two key deficiencies of the
tidal disruption model.  First, hydrocode simulations
\citep[e.g.,][]{Benz_Asphaug_1999} show that a high-velocity cometary
impactor of radius 10 to 20 km could disrupt a 200-km radius moon in
the region of the main rings, with the resulting fragments prevented
from re-accumulation by planetary tides.  Second, although
interactions with a gas disk are likely needed to bring the moon to
within the Roche limit, the event leading to the creation of the ring
can be delayed until a much later time when the gas disk had
dissipated, removing the vulnerability of the ring to loss via gas
drag.

The disruption of a ring progenitor moon requires a large number of
potential cometary impactors to be probable.  Estimates of the current
population of such objects are too low by at least an order of
magnitude to make such an event likely in the past billion years
\citep[e.g.][]{charnoz_lhb_2009}.  However, the number of outer Solar
System impactors could have been much larger in the distant
past. \citet{charnoz_lhb_2009} considered the Late Heavy Bombardment
(LHB) predicted by the so-called Nice model for the origin of the
structure of the outer Solar System.  They estimated near-certain
disruption of a hypothetical $\sim 200$--300 km radius progenitor moon
during this period, predicted to occur some 400 Myr to 1 Gyr after the
planets formed.  In the Nice model, the enhanced bombardment is driven
by the destabilization of a background planetesimal disk of comets as
Jupiter and Saturn cross a mutual mean-motion resonance and the orbits
of the ice giants are scattered outward \citep[see][for a review of
  the history of the cometary reservoirs, including a discussion of
  the Nice model and its successors]{dones2015}. In general, the
structure of the Kuiper Belt requires that Neptune migrated outward
via planetesimal scattering \citep[e.g.,][]{Malhotra_1995}, and this
implies both an initially more compact giant planet configuration and
a planetesimal disk containing between 10 and 100 Earth masses
\citep[e.g.,][]{Fernandez_Ip_1984,Hahn_Malhotra_1999}. The interaction
of the giant planets with such a massive disk would appear likely to
produce an early enhanced bombardment period even if the details of
the evolution differed from those of the Nice model.  Thus the
disruption of a Mimas-sized moon orbiting within the Roche limit
--\,if one existed\,-- seems probable during this period.

\begin{figure}
\figurebox{}{12pc}{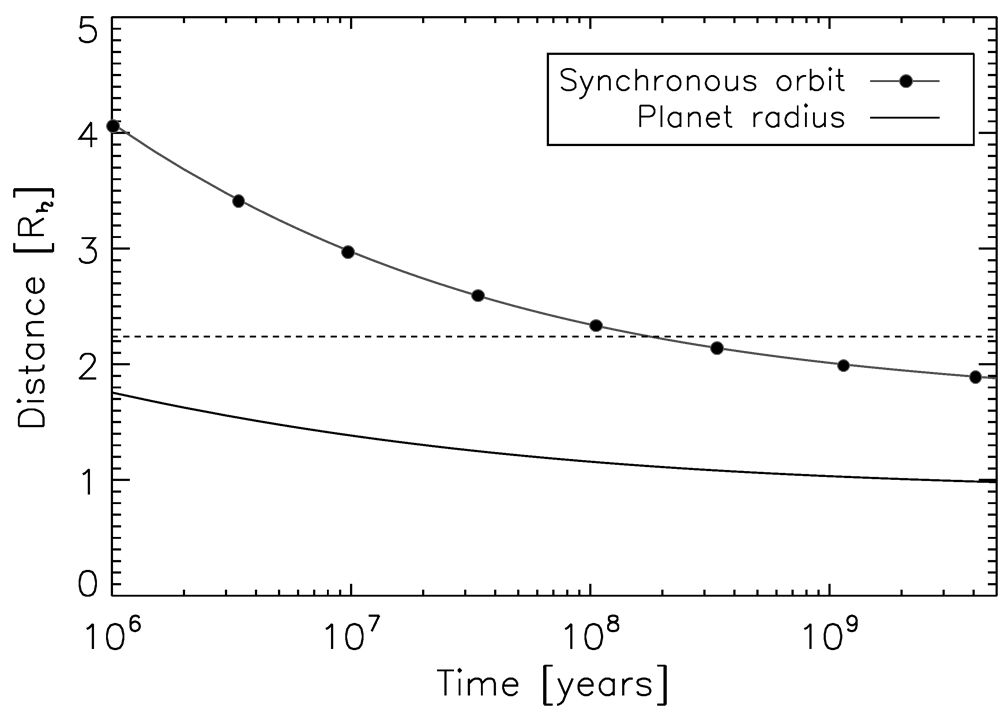}
\caption{Estimates for Saturn's radius (solid line) and the location of synchronous
  orbit (dotted solid line) for the first billion years of the planet's history
  \citep{Salmon_Canup_2016}.  The dashed line shows the approximate
  Roche limit for ice.}
\label{fig:async}
\end{figure}
	
Maintaining a ring progenitor moon interior to the Roche limit until
the time of the LHB places constraints on both the moon's mass and the
early tidal parameters for Saturn.  A moon exterior (interior) to the
synchronous orbit at Saturn migrates outward (inward) due to
interaction with tides it raises on the planet.  For Saturn's current
rotation rate, the synchronous radius is located at 112,000 km, or at
$a_{sync}\approx 1.92R_{S}$.  However, for the first billion years of
its history, Saturn would have been larger than its current size
\citep[e.g.,][]{Fortney_et_al_2007}. By conservation of spin angular
momentum, Saturn would have then rotated more slowly, with $a_{sync}$
initially well exterior to its current location, evolving inward to
the Roche limit after $\sim 10^8$ yr and to near its current location
in $\sim 10^9$ yr \citep[][and
  figure~\ref{fig:async}]{Canup_2010,Salmon_Canup_2016}.

A primordial ring progenitor moon that drifted inside the Roche limit
(e.g., via gas drag) would undergo inward tidal evolution, with rate
\begin{equation}
\frac {da}{dt} \sim \left (\frac{3k_2}{Q_p}\right ) \left (\frac
      {GM_p}{R_p} \right )^{1/2}\left ( \frac {M}{M_p}\right ) \left
      (\frac {a}{R_p} \right )^{-11/2}\ ,
\label{eq:tides}
\end{equation}
where $k_2$ and $Q_p$ are the tidal Love number and dissipation factor
for the planet, and $R_p$ is the early planet's radius (as distinct
from $R_{S}$, Saturn's current mean radius).  The timescale for a
primordial moon of mass $M$ to evolve tidally inward from the Roche
limit to the planet's surface is
\begin{align}
\Delta t &\sim
\frac{(r_R/R_p)^{13/2}}{20(k_2/Q_p)(M/M_p)(GM_p/R_p^3)^{1/2}}\cr &\sim
10^9 {\rm yr}\left (\frac{3\times10^{-6}}{k_2/Q_p}\right )\left
(\frac{5\times 10^{22} {\rm g}}{M}\right )\left
(\frac{1.1R_{S}}{R_p}\right )^5,
\label{eq:deltat}
\end{align}
where the second line considers an icy moon with density $\approx 1$ g
cm$^{-3}$ so that $r_R \approx 2.2R_{S}$, a somewhat enlarged planet
with $R_p = 1.1R_S$ (this quantity would actually be time-dependent as
the satellite evolved, as in figure~\ref{fig:async}), and $k_2/Q_p = 3
\times 10^{-6}$ for, e.g., $k_2 = 0.3$ and $Q_p = 10^5$.  Thus for a
primordial moon to survive within the Roche limit until the time of
the LHB requires both that the early $Q_p$ for Saturn was large
\citep[i.e., $\ge 3 \times 10^4$ for $k_2 = 0.3$;
  e.g.,][]{charnoz_chapter_2009} and that the moon was small, similar
in mass to the current rings.  In contrast, a moon that was, e.g.,
$10^2$ times more massive than the current rings would decay into
Saturn in $\le$ few $\times 10^7$ yr, even assuming a large $Q_p =
10^5$ (i.e., slow tidal evolution).  Similarly, even a small moon with
a mass a few $\times 10^{22}$ g would be lost if very rapid tidal
evolution applied to the early Saturn \citep[i.e., with $Q_p \sim
  10^3$, as has been advocated by some recent
  works:][]{Lainey_et_al_2012,Charnoz_2011,Lainey_2016}.
	
Assuming that an appropriate satellite could be delivered to and
maintained within the Roche-interior region until the time of the LHB,
a remaining question for the collisional disruption theory is that
disruption of a nominal rock-ice satellite would produce an initial
rock-ice ring, rather than a pure ice ring.  One possibility is that
the rock might be preferentially removed from the ring if it were
initially contained in much larger intact fragments that migrated
relative to the ice due to ring-moon interactions \citep[][also see
  section~\ref{subsec:MIR} below]{Charnoz_2011}. Alternatively,
disruption of an essentially pure ice progenitor moon would produce a
pure ice ring, but this then requires an explanation for the origin of
such an object.

\subsubsection{Tidal Disruption of a Comet Interloper}

\citet{Dones_91} proposed that Saturn's rings originated when a
heliocentrically orbiting Centaur (comet) of radius $\sim 200$ to 300
km passed well within Saturn's Roche limit and was tidally disrupted.
For an initially intact object on a parabolic encounter with the
planet, the inner portions of the object facing the planet upon
disruption will have a velocity somewhat less than the local escape
velocity from Saturn, and so will be weakly bound to the planet.  The
percentage of the interloper's mass that can be ``disintegratively
captured'' increases with the size of the interloper and with
decreasing periapse distance, reaching a maximum theoretical value of
50\% \citep{Dones_91,charnoz_chapter_2009}.  Disintegrative capture of
a rubble pile that was an intimate mixture of ice and rock would
produce a rock-ice ring.  However, if the object was instead
differentiated into a rock core and an icy mantle, the bound debris
could be overwhelmingly icy \citep{Dones_91}.  This has been
numerically confirmed in recent SPH simulations in the case of a
Titan-sized object, as we describe in the next section
\citep{Hyodo_2016_tidal}.  A differentiated state would be expected if
substantial melting of the comet's ice had occurred
\citep[e.g.,][]{Barr_Canup_2010}, suggestive of an intact object with
substantial strength.  How the inclusion of strength might modify the
disruptive capture process for comets comparable in mass to the
current rings is not clear; it would be unimportant for objects that
were much more massive.

Initially, captured debris would be on highly eccentric orbits with
semi-major axes of hundreds of saturnian radii
\citep{Hyodo_2016_tidal}.  In the absence of other processes,
dissipative collisions between debris particles would circularize
their orbits while approximately conserving angular momentum, leading
to a ring interior to the Roche limit \citep{Dones_91}.  However,
initial debris orbits would cross those of the regular Saturnian
satellites.  As such, the efficiency of ring production would be a
function of the rate of mutual debris collisions compared to the rate
of debris sweep-up by the satellites.  This process has not been
modeled.
	
For the current cometary flux, close passages by large comets appear
unlikely in the age of the Solar System \citep{Dones_91}. During an
outer Solar System LHB, such encounters would have been common,
although Saturn proves to be the giant planet least likely to have
experienced such an event
\citep{charnoz_lhb_2009,charnoz_chapter_2009}.  This is due to several
factors, including Saturn's low density, which makes the region over
which comet encounters lead to captured material smaller than those of
the other giant planets, which have higher densities
\citep{asphaug_benz_96}. Jupiter, Uranus and Neptune are estimated to
each receive about an order of magnitude more mass through close comet
encounters than does Saturn \citep[][their
  fig. 5]{charnoz_chapter_2009}. Thus, if Saturn's rings originated by
disintegrative capture, one would expect the other planets to have
ring systems at least as massive as Saturn's too.  The lack of such
systems suggests that while close passages of comets may have been
common in the early Solar System, their debris did not typically
produce long-lived rings, due to, e.g., escape of high-eccentricity
tidal debris from planetary orbit and/or efficient sweep-up of debris
by satellites \citep[e.g.][]{charnoz_chapter_2009}. An
  alternative solution would be that for some reason, the rings of
  Uranus and Neptune are short-lived while those of Saturn are
  long-lived. However, why there would be such a difference between
  the lifetime of rings of the different planets is not understood.
  This remains an open question.

\subsubsection{Recent Developments: The possibility
  of a super-massive primordial ring at Saturn}

While the mass of Saturn's current rings is many orders of magnitude
larger than that of the other ring systems, it still represents in
total only a small, several hundred km-sized moon.  The origin models
described above were designed to produce an initial ring whose mass is
comparable to that in Saturn's current rings, $\sim$ few $\times
10^{22}$~g. However, recent developments suggest that the current ring
mass may not actually reflect the initial ring mass. As described in
section~\ref{sec:proc}, the viscosity in a massive ring is
proportional to its mass squared, so that a very massive ring spreads
rapidly, causing its mass to decrease and its spreading rate to slow
until the spreading timescale becomes comparable to the age of the
system.  Simulations of the viscous evolution of a ring at Saturn have
shown that the current ring mass is comparable to that achieved as a
more massive initial ring viscously evolves over 4.5 billion years
\citep[figure~\ref{fig:ringmass};][]{Salmon-etal-2010}, independent of
the starting ring mass. This agreement could, of course, be
coincidental.  However, perhaps a more compelling interpretation is
that the mass of Saturn's current rings reflects their dynamical age,
which could be billions of years. Other data would be necessary to
unambiguously infer the rings' using radiogenic dating
technics. Ideally a sample return mission from Saturn's rings would be
the best, but does not seem within reach of current technologies.  The
primordial progenitor of Saturn's ring system then could have been
much more massive and still, after billions of years of collisional
evolution, have left a ring comparable in mass to that seen today.

How could such an initially massive ring have formed? Based on the
arguments above, formation by collisional disruption of a pre-existing
satellite does not appear likely, as it would be extremely improbable
in the short time a large moon would spend inside the Roche limit
before tidally evolving into the planet. For example, per
Eq.~(\ref{eq:deltat}), a $10^{25}$~g satellite would tidally decay
from the Roche limit to Saturn's surface in only a few million years
even for large $Q_p$.

In contrast, tidal disruption could produce a massive ring if the
interloper was massive, in a scaled-up version of the \citet{Dones_91}
model.  This possibility has been recently considered by
\citet{Hyodo_2016_tidal}, who consider tidal disruption of a single
hypothetical differentiated Titan-sized passing Kuiper Belt
Object.\footnote{No KBOs of this size have yet been discovered. The
  most massive Kuiper Belt Object known, Eris, has a mass $12\%$ that
  of Titan. However, much more massive KBOs may have been present in
  the early Solar System and may yet exist in the distant Kuiper Belt
  or inner Oort Cloud.} \citet{charnoz_lhb_2009} predict that close
passes by $500$~km $\le R\le 2000$~km objects would be rare events
even during the LHB, but the probability of such an event depends
sensitively on the size distribution of large objects at that time,
which is uncertain. Using hydrodynamical simulations,
\citet{Hyodo_2016_tidal} find that a differentiated Titan-sized object
passing as close as 3 planetary radii may have its icy mantle
shattered by tides, producing an ice-rich debris disk whose mass is
compatible with the formation of Saturn's current rings, as well some
or all of Saturn's inner moons (see section~\ref{subsec:MIR}
below). The silicate-rich core, as well as most of the
  incoming object's mass, go onto hyperbolic orbits and are lost from
  the planet. However, a very small portion of the silicates (a
  fraction of a percent) may also be captured and may contribute to
  the current silicate content of the moons. An appealing aspect of
  this scenario is that it can produce an ice-dominated ring at Saturn
  with little silicates. However, tidal disruption in general would
be equally likely to produce a prograde or a retrograde ring, as the
final debris would orbit in the same direction as the passing body.
But clearly a single prograde system like that at Saturn would result
half the time.

A more challenging issue is that a size distribution of background
objects consistent with such an event at Saturn would imply that the
other outer planets would have had massive rings produced by tidal
disruption too, as discussed in the prior section and in
\citet{charnoz_lhb_2009}. Such rings would have viscously evolved to
an asymptotic state whose mass is many orders of magnitude larger than
those of the other very low-mass ring systems, per the arguments in
section~\ref{subsec:Spreading} and in figure~\ref{fig:ringmass}. An
additional process would then be needed to remove such early massive
rings at the other giant planets but not at Saturn.  It is not clear
what could accomplish this.  Processes associated with the Sun's
radiation can remove small particles through inward orbital decay, but
these would be substantially weaker at Uranus and Neptune than at
Saturn for a given particle size.  We return to the interesting issue
of potential early massive rings at Uranus and Neptune in
section~\ref{subsec:UN}.

Alternatively, a massive ring could be produced through tidal
disruption if a massive satellite migrated interior to the Roche
limit.  It has long been recognized that interaction with a primordial
gas-rich disk around Saturn would cause small moons to spiral inward
due to aerodynamic gas drag \citep[e.g.,][]{Harris_1984}. The
migration rate due to this process is inversely proportional to the
moon's physical radius, and is thus not important for large moons on
relevant timescales. More recent work has focused on the ability of
density wave interactions with a gas disk to modify the orbits of
large satellites
\citep[e.g.,][]{Canup_Ward_2002,Mosqueira_Estrada_2003a,Mosqueira_Estrada_2003b}. Such interactions can cause a satellite's orbit to spiral inward with a
rate that is proportional to the satellite's mass \citep[Type I
  migration;][]{ward1986}, or, for even larger satellites capable of
opening gaps in the gas disk, the satellite's orbital motion becomes
coupled to the local viscous expansion of the disk \citep[Type II
  migration;][]{Lin_Papaloizou_1986}.  The survival of Titan and
galilean-sized moons against such potentially destructive processes
has provided new constraints on satellite formation conditions and
motivated the development of new satellite origin models
\citep[e.g.][]{Canup_Ward_2002,Mosqueira_Estrada_2003a,Mosqueira_Estrada_2003b,Alibert_et_al_2005}.

An aspect on which all models do not agree, and that is still
  an open question, is the delivery of solids to the circumplanetary
  disk. Indeed, as grains grow and progressively decouple from the gas
  in the outer Solar System, it is increasingly difficult for them to
  penetrate a planet's circumplanetary disk due to the pressure
  gradient created by the planet in its surroundings. While
  \citet{Canup_Ward_2002} consider that grains are small and are
  transported with the gas, in
  \citet{Mosqueira_Estrada_2003a,Mosqueira_Estrada_2003b} and
  \citet{Alibert_et_al_2005}, grains are considered to be big and
  decoupled. They are scattered in the disk by nearby planets. Recent
  works modeling ``pebble accretion'' advocate an essentially bi-modal
  planetesimal population in which small pebbles are accompanied by $>
  100$-km planetesimals formed via gravitational instability (e.g.,
  \citet{levison2015}), which could imply a larger population of
  small, gas-coupled objects than assumed by prior works that consider
  power-law planetesimal size distributions.  Overall, the mechanism
  of solid delivery remains an open and important question, which
  depends strongly on the relative timing of grain growth and giant
  planet growth in the outer Solar System.
 
While models for the formation of outer planet satellites vary, a
general implication of density wave interactions is that they provide
a means for much more massive satellites to migrate inward to the
Roche-interior region, where they might then become a source for ring
material.

Perhaps the most explored model to date is one in which regular
satellites form within a disk supplied by an ongoing inflow of gas and
solids from heliocentric orbit as a gas planet completes its growth
\citep{Canup_Ward_2002,Canup_Ward_2006,Alibert_et_al_2005,ward2010,Sasaki_et_al_2010,Ogihara_Ida_2012}.
These models are based on the recognition that the period over which
inflow to the disk occurs may be long, comparable to the lifetime of
the solar nebula ($\ge 10^6$ yr), so that satellites may
  accrete during the inflow phase, rather than after it, as assumed
by prior models.

In an actively-supplied disk, the gas component likely reflects a
quasi-steady state between the inflow supply and viscous spreading, so
that as the nebula dissipates and the inflow slows, the disk becomes
increasingly ``gas-starved''.  Solids flowing into the disk provide
the source material for growing satellites, while Type I interaction
with the gas disk cause each satellite's orbit to spiral inward at a
rate proportional to its growing mass.  The balance of these two
processes causes there to be a critical maximum mass for a satellite
of a gas giant planet, which for reasonable disk and inflow parameters
is comparable to the mass of Titan at Saturn, and the mass of the
galilean satellites at Jupiter \citep{Canup_Ward_2006}. Each satellite
grows no larger than this critical mass before it spirals inward into
the planet.  As satellites are lost, new ones grow in their place as
more solids flow into the disk.  Multiple generations of large
satellites form and are lost, each having a similar mass
  compared to that of the host planet, with the satellite system mass
  oscillating about a quasi-steady-state value of $10^{-4}$ planet
  masses, independent of the total mass processed through the disk
  \citep{Canup_Ward_2006}.  This value is comparable to the satellite
  system mass ratios observed around all the outer planets.  The
  overall process continues until the inflow itself ends and the last
system of satellites stabilizes as the gas disk dissipates and Type I
migration ends
\citep{Canup_Ward_2002,Canup_Ward_2006,Sasaki_et_al_2010,Ogihara_Ida_2012}.
Galilean-like systems with multiple large satellites are the most
common outcome seen in direct $N-$body simulations of this process
\citep{Canup_Ward_2006,Ogihara_Ida_2012}, but systems with one large
Titan-like satellite are possible if inner large satellites migrate
inward and are lost as inflow to the planet ends \citep[][her
  Fig.~1]{Canup_Ward_2006,Canup_2010}.
  
Other works consider that the satellites formed after the
  inflow to the disk ended, so that the disk has a constant total
  mass, and envision a very low-viscosity disk so that large
  satellites may open gaps and halt their orbital migration
  \citep{Mosqueira_Estrada_2003a,Mosqueira_Estrada_2003b}.  In this
  case, satellite survival is predicted for satellites that exceed a
  critical gap-opening mass, implying that satellites of this mass or
  greater may survive.  For an inviscid disk, the gap-opening mass can
  be comparable to the mass of Titan and the Galilean satellites
  \citep{Mosqueira_Estrada_2003a,Mosqueira_Estrada_2003b}, and for
  this case the constant-mass disk model predicts a minimum satellite
  system mass ratio of order $10^{-4}$ planet masses.  However, in
  this model, there is not a limit on how much larger than the
  gap-opening mass satellites may grow, and instead, the final mass of
  the satellites is a function of the assumed initial disk mass.  This
  conceptually distinguishes the constant-mass disk models from the
  actively-supplied disk models, because in the former the final
  satellite system mass depends on the assumed mass of the initial
  disk (which is highly uncertain), while the latter predict a common
  satellite system mass ratio independent of the mass processed
  through the disk.

In both the actively-supplied disk and constant-mass disk
  models, inward migration of satellites to within the Roche limit may
  occur due to density wave interactions with the gas disk.  In the
  inflow-supplied disk models, Type I migration leads to repeated
  losses of Titan-sized objects at Saturn.  Smaller inner satellites
  might evolve inward due to tides, but satellites migrate inward
  over large distances due to density wave interactions only once they
  reach very large, Titan-like masses \citep{Canup_Ward_2006}.  In
  constant-mass, inviscid disk models, large satellites would be
  protected from migration by opening gaps, while smaller satellites
  might migrate inward due to Type I migration and be lost.  The
  latter merits further consideration by direct models of satellite
  accretion within such disks to determine the properties of potential
  ring progenitors.

\subsubsection{Tidal Stripping from a Large Satellite}
\label{sec:TSLS}

The developments above --\,in particular, the prediction that one or
more Titan-sized satellites at Saturn migrated into the planet at the
end of the satellite formation era\,-- motivated a new model for ring
origin, in which a massive primordial ring is formed as tides strip
the icy outer layers from a large satellite as it spirals into Saturn
\citep{Canup_2010}.

For a gravity-dominated, Titan-sized satellite, tides will begin to
remove mass once the satellite's orbit migrates within the Roche limit
set by its mean density, which is located at $\approx 1.75\,R_{S}$ for
a Titan-density satellite. The nature of the resulting mass loss
depends on the satellite's interior structure.  An undifferentiated,
uniform composition satellite would disrupt completely.  However, a
Titan-sized moon would be expected to have a differentiated interior,
with an ice mantle overlying a rocky core, due both to the energy of
its accretion and strong tidal heating as its orbit spiraled toward
the planet; see detailed calculations and estimates in
\citet{Barr_Canup_2010} and \citet{Canup_2010}.  For a differentiated
ice-rock satellite, tides will first remove material from its outer
ice shell.  The preferential removal of ice then causes the
satellite's mean density to increase slightly until the remnant
satellite is marginally stable at its current distance
\citep{Canup_2010}. Continued inward migration (due to, e.g., Type I
migration and tidal interaction with Saturn) would then lead to
additional ice removal, with the process continuing until either the
remnant satellite collides with the planet or its higher-density rocky
core disrupts \citep{Canup_2010}. The Roche limit for rock of density
$\rho_{\rm rock}$ is at $r_{R,\rm rock} = 1.5\,R_{S}(3 \;{\rm g\;
  cm^{-3}}/\rho_{\rm rock})^{1/3}$. Planet contraction models predict
that Saturn's early radius would have been between $R_p = 1.5\,R_{S}$
and $R_p = 1.7\,R_{S}$ \citep[][see also figure~\ref{fig:async}
  above]{Marley_et_al_2007,Fortney_et_al_2007}.  For $r_{R,\rm rock}
\le R_p$, the remnant satellite would collide with the planet before
its rocky core disrupts.  In this case tidal stripping would produce
an essentially pure ice ring. The silicates of today's
  satellites may have been provided by a tiny fraction of silicates
  stripped from the initial Titan-sized progenitor \citep[see][and
      section~\ref{subsec:MIR}]{Charnoz_2011}.

\begin{figure}
\figurebox{}{10pc}{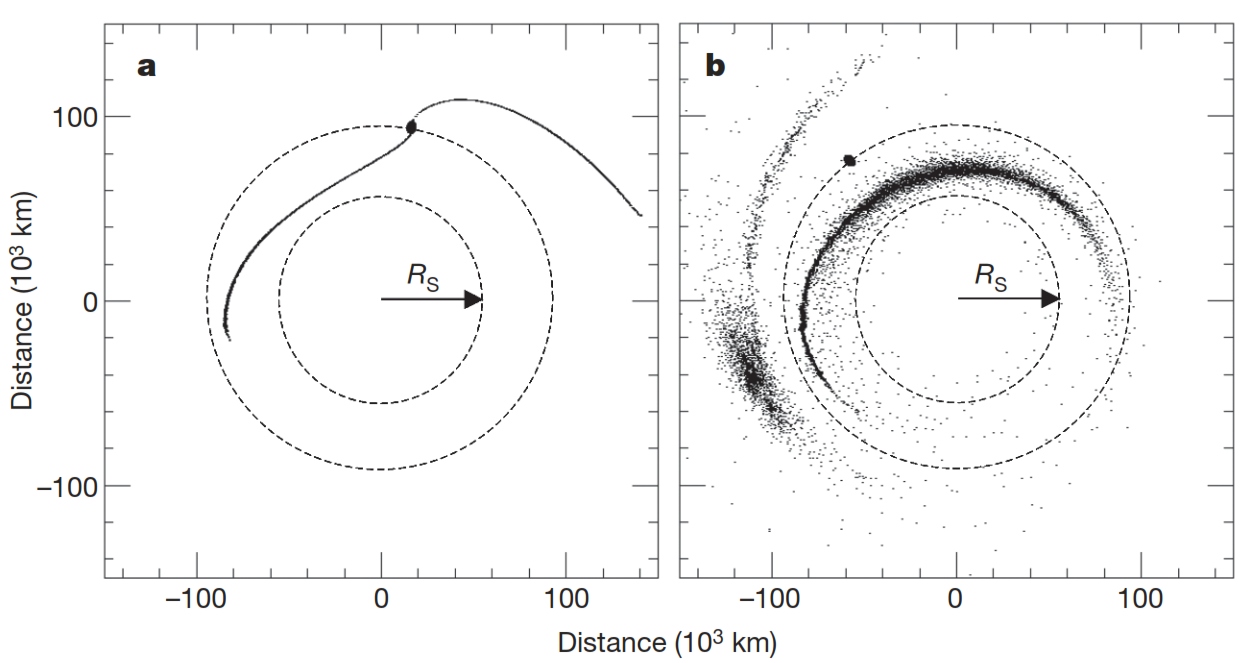}
\caption{SPH simulation showing the tidal removal of ice from a
  differentiated, Titan-mass satellite, from \citet{Canup_2010}.  Type
  I migration and tidal interaction with the planet cause the
  satellite's orbit to spiral inward from its initial Roche limit
  ($a_{max}$) to the planet's surface ($R_p$) in $\sim 10^4$ yr.  The
  satellite's evolution across this region is tracked with a series of
  SPH simulations that treat the satellite and planet
    explicitly but do not include the gas disk.  The satellite starts
  with $a = a_{max}$ and is evolved for several orbits with SPH to
  simulate tidal mass removal and the establishment of a stable
  satellite remnant. The remnant satellite is then shifted inward by
  $\Delta a \sim 10^{-2}a$ and re-simulated, with the process repeated
  until $a$ is small enough that the satellite's rocky core disrupts,
  which determines the minimum planetary radius consistent with the
  creation of a pure ice ring. Frames here show tidal stripping from a
  50\% ice, 50\% rock satellite at $a = 0.97a_{max}$ after 8 simulated
  hours [a] and 25 hours [b]. Distances are shown in units of $10^3$
  km; for comparison, Saturn's B and A rings lie between $\sim$ 92,000
  and 137,000 km. Dashed circles indicate the satellite's orbit and
  Saturn's current mean radius, $R_{S}$; Saturn's radius at the time
  of the satellite's decay was likely $R_p \ge 1.5R_{S}$ (e.g.,
  figure~\ref{fig:async}).}
\label{fig:stripping}
\end{figure}

Figure~\ref{fig:stripping} shows a hydrodynamical simulation to
simulate tidal stripping from a differentiated Titan-sized satellite
as it spirals inward from its initial Roche limit \citep{Canup_2010}.
Material originating from the satellite's ice mantle is lost through
its inner and outer Lagrange points (L$_1$ and L$_2$), leading to
particles on eccentric orbits (with $e \sim 10^{-1}$) with semi-major
axes interior and exterior to that of the satellite, respectively.
Subsequent collisions between particles will tend to circularize their
orbits.  Interior particles may collide directly with the planet or be
driven into the planet by the satellite as it continues to migrate
inward, while exterior particles can supply the eventual ring. The
total mass of ice produced via tidal stripping depends on the
satellite mass and the location of the planet's surface at the time of
the satellite's demise.  For a Titan-sized body, in the limit that
$a_{R,rock} = R_p$, $\sim 10^{25}$ g of ice is stripped into orbits
exterior to the satellite for a Titan-sized body ($\sim 10\%$ of the
original satellite's mass), while if $a_{R,rock} < R_P$, less ice is
produced before the remnant satellite hits the planet
\citep{Canup_2010}.

As each ringlet of ice is stripped, strong shepherding torques from
the remnant satellite rapidly repel it, driving exterior material into
orbits beyond the Roche limit for ice in $\sim 10^2$ yr. Once this
material orbits substantially beyond its Roche limit \citep[e.g. $>
  1.1 r_{R,ice}$;][]{Canup_Esposito_1995,Canup_Esposito_1996} it can
rapidly accumulate, forming a pure ice moon whose final mass is $m_o
\le 10^{25}$ g.  The ice stripped from the progenitor satellite is
then stored for a time in this secondary moon, which on a longer
timescale (because its mass is substantially smaller than the original
Titan-sized progenitor) spirals back inward due to tidal interaction
with Saturn.  The ice moon takes $\sim 10^6$ yr $\left[3\times
  10^{-6}/(k_2/Q_p)\right] \left[10^{25} g/m_o\right]$ to tidally
decay back within the Roche limit, where it disrupts into a massive
ice ring \citep{Canup_2010}. This expression considers a small value
for $k_2/Q$ for primordial Saturn, based on traditional estimates
(e.g., \citet{goldreich_soter_1966}).  This is not necessarily
inconsistent with much more rapid tides observed today (e.g.,
\citet{Lainey_et_al_2012,Lainey_2016}), as recent work proposes that
the current values reflect resonance locking between moons and
internal oscillation modes in the planet, a process that implies
larger effective $Q$ values in the past \citep{fuller2016}.  We return
to this issue in section~\ref{subsec:cuk}.

Because the tidal stripping model relies on interaction with the gas
disk (e.g., Type I migration) to deliver the ring progenitor to the
Roche interior region, the potential vulnerability of the ring to loss
via gas drag must be considered \citep{Canup_2010}.  Consider a gas
inflow to Saturn that decays exponentially with time with a time
constant $\tau_g \sim 10^6$ yr, reflecting a waning inflow to the disk
due to the dispersal of the solar nebula.  That Titan survived while a
similarly massive interior satellite was lost implies that Titan's
timescale for Type I decay was comparable to $\tau_g$, implying a very
low $\Sigma_g \sim $ few to 10 g\,cm$^{-2}$ at the time the last large
inner satellite was lost \citep{Canup_2010}. The gas density would
decrease further in the $\sim 10^6$ yr required for the secondary moon
to tidally decay inward and be disrupted.

Thus a ring produced by tidal stripping can survive gas drag until
after the gas disk has dissipated, so long as the progenitor
  satellite migrates within the Roche limit near the very end of the
  gas disk lifetime.  Survival of the ring thus does require special
  timing: the actively-supplied disk models show that many large
  satellites were likely lost to inward migration, and the rings
  produced by nearly all of them would have been lost to gas drag.
  But at some point we know that the gas disk around the planet
  dissipated and the loss of large satellites via inward migration
  must have ended.  The key question is then: what was the fate of the
  last ring created by such a process?  For Saturn, the lack of a
  large inner companion to Titan (one of the most challenging features
  to explain in the Saturnian system) may imply the very late orbital
  decay of the inner body, and this implies conditions favorable for
  ring survival as described in the prior paragraph.  However even the
  final episode of tidal stripping could have been unsuccessful in
  producing a long-lived ring in other circumstances.  For example in
  the context of the actively-supplied disk models, Jupiter, in
  contrast to Saturn, must have retained its final generation of large
  interior satellites (i.e., Io and Europa) as inflow to that planet
  ended.  One expects that Jupiter would still have lost large
  interior satellites from prior generations of moons that formed when
  the gas inflow rate to the planet was higher and the disk was more
  gas-rich \citep{Canup_Ward_2002,Canup_Ward_2006}.  Jupiter's final
  ring produced by tidal stripping would have likely been lost to
  stronger gas drag.  Thus tidal stripping offers a mechanism that
  would frequently have produced rings as large satellites migrated
  within the Roche limit, but only the final such rings could survive,
  and even then not in all circumstances.  This offers an appealing
  explanation for why Saturn alone has a massive ring today.

An overall strength of the tidal stripping model is that it would
naturally produce an essentially pure ice, prograde ring. While the
model was developed under the premise of a particular model of
satellite formation \citep{Canup_Ward_2002,Canup_Ward_2006},
it could apply to other situations and other models as well.
The basic requirements for the production of a long-lived ring via
tidal stripping model are (1) a differentiated rock-ice satellite that
orbitally migrates within the Roche limit, and (2) a ring decay
timescale due to gas drag (Eq.~\ref{eq:gd}\,) that exceeds the
lifetime of the gas disk. The case for a differentiated ice-rock
satellite is strongest for a very large satellite because of its
substantial accretional and tidal heating, which would generally imply
a massive ring even if other aspects of the evolution or the satellite
accretion model differed.

  An important overall caveat to the tidal stripping model is
  that it considers ring formation as a byproduct of satellite
  formation, a broad topic which itself remains quite uncertain.
  Multiple key general uncertainties persist in our understanding of
  the protosatellite environment, including the radial and temporal
  structure of the circumplanetary disk, the time variation of the
  inflow of gas to such a disk, the presence or not of viscosity and
  the accretion rate onto the planet, the delivery of solids to the
  disk and their accretion within the disk, and the role that the
  magnetic field may or may not play in providing turbulence.
  Substantial progress may result from future hydrodynamical
  simulations of giant planet growth that simultaneously consider
  inflowing material, as well as the accompanying disk and planet
  structure, although it is clear that such models are still
  challenging for even high-performance simulations and will likely be
  for years to come.  It may also be possible to apply recent advances
  in planetesimal formation made in the context of planet accretion
  (e.g., \citet{levison2015} and references therein) to make better
  progress in understanding satellite growth as well.  Improved
  knowledge of these many processes will be necessary to better
  evaluate our models of satellite formation, as well as the ring
  origin models linked to the satellite formation environment,
  including the condensation and tidal stripping models.

\subsubsection{Implications of a Massive Initial Ring}
\label{subsec:MIR}

A massive primordial Saturnian ring requires an explanation for what
happened to perhaps $\sim 99\%$ of the ring material as it evolved
over 4.5 billion years.  As a ring viscously spreads, ring material is
depleted both by collision onto Saturn at the ring's inner edge, and
by spreading of material beyond the Roche limit at the ring's outer
edge. Once ring material is substantially outside the Roche limit,
ring material can accrete into satellites. Spawned moons evolve
outward, due, at first, to resonant interactions with the ring and
then by tidal evolution, with the mass of spawned moons decreasing
with time as the mass of the ring decreases due to viscous spreading
\citep{Crida_Charnoz_2012}. This process was first directly simulated
for a ring similar in mass to Saturn's current rings by
\citet{Charnoz_2010}, who demonstrated that the very small innermost
Saturnian moons (out to and including Janus) were likely spawned from
the rings in the recent past.  However, it was clear based on that
work that a more massive ring would give rise to more massive moons as it
viscously evolved.

\citet{Canup_2010} estimated that a massive ring produced via tidal
stripping from a Titan-sized satellite would have a total mass and
angular momentum consistent with the current rings and the inner
Saturnian moons out to and including Tethys ($\approx 5R_{S}$), and
that the first spawned satellite from such a ring would have a mass
comparable to that of Tethys ($\approx 16 \times$ the mass of Mimas).
These expectations have generally been confirmed by numerical
simulations \citep{Charnoz_2011,Salmon_Canup_2015DPS}. However, the
final orbital architecture of the satellite system spawned from a
massive ring depends on the assumed value of the tidal parameters for
Saturn. Estimates of the time-averaged tidal parameter for Saturn have
traditionally been in the range $10^{-6} \le k_2/Q_p \le 10^{-5}$
\citep{goldreich_soter_1966}, which implies that the moons out to
Tethys (or their progenitors, since the inner moons may have been
disrupted and re-accreted \citep{charnoz_chapter_2009}) were spawned
from the rings \citep{Salmon_Canup_2015DPS}.  This offers an appealing
explanation for the unusually ice-rich composition of these inner
moons, which, as a group, are about 90\% ice by mass.  Tethys
  in particular is the most ice-rich satellite larger than 100~km in
  radius known in the Solar System, containing $\le 6\%$ rock by mass.
  In contrast, the large Uranian satellites contain about half rock,
  and even low-density Miranda contains about $20\%$ rock. However,
Enceladus is currently about half rock, and if it or its progenitor
were spawned from an icy ring, this rock would need to have been
somehow supplied by external sources.  One possibility is that later
impacts delivered rock to these moons.  Indeed, the estimated rock
delivered to the inner Saturnian moons during the LHB is roughly
comparable to their current total rock content
\citep{canup2013,Salmon_Canup_2014DPS}. However, whether external
bombardment alone can explain the current distribution and variation
of densities for the inner moons is not clear, and remains an
outstanding issue for such models. Alternatively, there may
  have been a small portion of rock in the initial ring due to, e.g.,
  tidal stripping from an incompletely differentiated satellite, or
  tidal disruption of a large body that leaves some portion of the
  object's silicates in orbit as well.

Satellites spawned from a massive ring would have achieved orbits well
beyond that of Tethys if the outward tidal evolution of satellites had
been much more rapid.  Recent astrometric results imply $Q_p\sim
10^3$, or $10^{-4} \le k_2/Q_p \le 10^{-3}$ for the current Saturn
system \citep{Lainey_et_al_2012,Lainey_2016}.  The evolution of a
massive ring and its spawned moons was independently simulated in
\citet{Charnoz_2011}, who considered rapid tidal evolution, with
$k_2/Q_p \sim 10^{-4}$ motivated by these results.  For such rapid
tides, a massive ring could spawn not only the inner Saturnian
satellites, but objects as distant and massive as Dione and Rhea as
well \citep[see also][]{Crida_Charnoz_2012}. Dione and Rhea contain
$\sim 10^{24}$ g in rock, much more than the total rock in the inner
moons ($\sim 10^{23}$ g). \citet{Charnoz_2011} argued that Saturn's
initial ring had a substantial rock component, with the rock initially
in the form of large chunks that were able to open gaps in the ring
and undergo Type II migration, so that the ring's rock was
preferentially removed.  Using a direct simulation of a Rhea-sized
rocky chunk embedded in a ring, they showed that the chunk is expelled
from the ring's outer edge, where it could then form the core of a
spawned satellite.  They suggested that stochastic variation in such a
process could explain the varied densities of the inner saturnian
moons.  Whether this rock removal process would be efficient enough to
yield a nearly pure ice ring for a realistic initial size distribution
of rock fragments is not clear.  Further, the
\citet{Lainey_et_al_2012} determination of $Q_p \sim 10^3$ for
Saturn, while intriguing, remains controversial because it implies a
much different $Q_p$ for Saturn than similar techniques indicated for
Jupiter \citep{lainey2009}. However, a recent re-evaluation of $Q_p$
based on astrometric data and using Cassini images was independently
obtained by two teams (IMCCE in Paris and JPL in USA) using different
tools. Both teams agree that $Q_p$ is indeed very small, so that
Saturn's tides appear strong, at least for the current system
\citep{Lainey_2016}.  Whether this low value has applied to Saturn
over its entire history is not known.

Thus substantial progress has been made on the possible formation of a
massive ice ring at Saturn as the precursor to the current rings.
However, there remain important open issues concerning whether and how
such a ring could evolve into the system of rings and moons we see
today.

\subsubsection{Pollution of an Early-Formed Saturnian Ring}

Explaining how Saturn's rings remain so rock-poor today presents an
ongoing challenge to all of the origin models we have discussed, which
invoke either a primordial origin, or an origin some 1~Gyr later
during the LHB.  Thus all imply a ring age of several
billion years.  Even if one begins with a pure ice ring, prior
estimates suggest that the ring would excessively darken over $\sim
10^9$ yr.  To avoid this, the pollution rate due to micrometeoroid
bombardment estimated previously \citep[e.g.,][]{Cuzzi_Estrada_1998}
must be too rapid, due to either an underestimation of the rings'
total mass or to an overestimation of the bombardment rate, or both
\citep[e.g.,][]{Cuzzi_et_al_2010}.  In addition, some Cassini
observations and $N$-body simulations have been interpreted to suggest that the B ring contains
substantially more mass than Voyager-era estimates
\citep{Robbins_et_al_2010}, implying a higher likelihood that the
rings are primordial, although others infer a total mass similar to
prior estimates \citep[e.g.][]{Reffet_et_al_2015} or somewhat lower
\citep{hedman2016}. Whereas all estimates are about the same order of
magnitude, better constraints are expected from the end of the Cassini
mission, in particular from the first direct measurement of the total
ring mass as the spacecraft undergoes close passes to the rings prior
to its descent into Saturn.

\subsubsection{A model for making young rings : resonant collisions in the satellite system}
\label{subsec:cuk}

\citet{cuk2016} propose that recent collisions between icy
  moons could provide a viable mechanism to form today's rings. Their
  argument is two-fold, and is based on the recent measurements of
  Saturn's tidal dissipation, the so-called $Q_p$ parameter
  present in Eq.~(\ref{eq:tides}), by \citet{Lainey_2016}. The
  recently determined value of $Q_p$ implies intense tidal dissipation
  inside Saturn and is consistent with the heat flux at the surface of
  Enceladus. As a consequence, Saturn's satellites may undergo a rapid
  tidal evolution of their orbits. If Saturn's satellite system was as
  old as Saturn itself, then this would imply that Tethys and Dione
  may have crossed their mutual 3:2 resonance in the recent past
  (about 100 Myr ago). During this resonant configuration, both orbits
  become significantly eccentric and inclined. While the
  eccentricities can damp due to eccentricity tides, the inclinations
  of Saturn's mid-sized moons are not expected to evolve appreciably
  over the age of the Solar System. The final inclination of Dione is
  always substantial, typically comparable to a degree, while
  the observed inclination of Dione is only $0.028^\circ$.  So the
  authors conclude that Tethys and Dione have never crossed this
  mutual resonance, and that one way to solve this paradox would be
  that they are younger. Thus they propose that a ``proto-Rhea'' would
  have crossed a resonance with ``proto-Dione'' about 100 Myr ago. The
  proto-Rhea may have been already been placed on an eccentric orbit due to an
  evection resonance with the Sun.  Proto-Rhea and proto-Dione could
  have collided with a velocity of about 3 km/s, enough to
  catastrophically disrupt both satellites, and releasing enough
  material to create a new generation of satellites, plus the
  rings. This scenario, however, has not been fully tested
  numerically. Several problems may prevent the debris disk from
  forming today's rings. Whereas the spreading timescale of the debris
  down to the Roche limit may be as fast as 1000 years, the concurrent
  reaccretion into satellites may act on a similar timescale, leading
  to a stopping of the spreading process and feeding of the Roche
  Limit. Also, how the debris could spread all the way to inside
    $r_R$, across the orbits of Enceladus and Mimas, is unclear, as is
    how the process would produce rings that are overwhelmingly icy.
  So, while it is clear that a large part of the debris disk would
  reaccrete into satellites, it is not clear if a small fraction would
  form Saturn's rings before being accreted by the satellites. This
  interesting scenario must be studied in more detail in the
  future. Beyond the specific case of Dione and Rhea, the
  \citet{cuk2016} paper emphasizes the possible role of evection
  resonances in Saturn's satellite system that may regularly
  destabilize the system, and possibly lead to the destruction and
  re-accretion of several generations of satellites over the age of
  the Solar System.

\subsubsection{What would we need to make further progress on the origin of Saturn's rings?}

We have mentioned different possible scenarios for the origin of
Saturn's rings. The main problem to compare them on an equal ground is
that they do not have all the same level of maturity. Some scenarios
for ring and satellite origin have been studied in different papers,
mixing numerical simulations and semi-analytical models, while others
are still based only on qualitative arguments (like the young rings
model). In addition, the structure of a circumplanetary disk is still
not well constrained. So the first thing we would need to make further
progress is a modeling effort on the least explored aspects, including
the structure of the circumplanetary disk, as well as a better
understanding of the giant planet formation process in the context of
the early Solar System. Then we would need new data. Today it is still
difficult to design a critical observable that would help to
distinguish between these models. Clearly a precise determination of
the micro-meteoroid flux at Saturn would help to give surface ages for
the rings and satellites. We await the publication of the dust
  flux onto the rings by Cassini's CDA team.  The ``grail'' would be a sample
return from Saturn's rings, but that may not happen for many years.

\section{Rings of Jupiter, Uranus and Neptune}
\label{sec:JUN}

\subsection{Rings of Jupiter}

The jovian rings are the only known example of a system comprised
solely of ``ethereal'' rings \citep{burns1984}. That is, while Jupiter
has distinct main and halo rings and two ``gossamer'' rings, all of
these components have optical depths $\tau \ll 1$, so that the rings
do not present enough surface area to be detected by occultation
experiments, but only in images, most easily when the rings are viewed
nearly edge-on. Saturn, Uranus, and Neptune all have ethereal rings
too, but in addition have rings of higher $\tau$.  Ethereal rings
consist largely, or in some cases almost entirely, of ``dust,''
particles with sizes in the range of 0.1--100~$\mu m$. Dust particles
are subject to non-gravitational forces
\citep{burns1984,Burns_et_al_2001} such as Poynting-Robertson
  drag, radiation pressure and electromagnetic forces.

Jupiter has four known moons within the orbit of the innermost
galilean satellite, Io -- moving outward, Metis, Adrastrea, Amalthea,
and Thebe. The two gossamer rings, with optical depths of order
$10^{-7}$ and $10^{-8}$, are clearly associated with Amalthea and
Thebe because the rings' outer edges coincide with the orbits of the
two moons. In addition, the thicknesses of the gossamer rings match
the vertical excursions of Amalthea and Thebe from Jupiter's
equatorial plane (thousands of km) due to their slightly inclined
orbits. The gossamer rings have been interpreted as dust liberated
from Amalthea and Thebe by micrometeoroid impacts that evolves inward
under Poynting-Robertson (P-R) drag
\citep{burns1999,showalter2007}. For micron-sized particles, the P-R
drag timescale is $\approx 10^5$~years \citep{burns1984,burns2004}.

Moving inward from the gossamer rings, the ``main'' jovian ring is
roughly bounded by Metis and Adrastea, which are much smaller than
Amalthea and Thebe. The optical depths in macroscopic particles
$\tau_L$ and in dust $\tau_S$ in the main ring are comparable. For
instance, \citet{throop2004} find $\tau_L \approx 4.7 \times 10^{-6}$
and $\tau_S \approx 1.3 \times 10^{-6}$. Metis and Adrastea are
important sources for the main ring, but unlike the gossamer ring,
additional source bodies must be present. \citet{burns2004} estimate
that the two moons ``comprise only about one-third of the [main]
ring's total area in source bodies.'' Metis and Adrastea have much
smaller orbital inclinations than Amalthea and Thebe, and the main
ring is physically much thinner (of order 100~km), compared with the
gossamer rings.

Finally, the innermost jovian ring is the ``halo,'' which has a wide
radial extent ($\approx$~30,000~km), with its outer edge near the
inner edge of the main ring. The halo is physically thick; although
most of the particles in the halo lie within a few hundred km of
Jupiter's equatorial plane, the halo is still detectible 10,000~km
from the plane. The radial extent of the halo is bounded approximately
by the 2:1 Lorentz resonance and the 3:2 Lorentz resonance at 1.41 and
1.71 Jupiter radii, respectively. Lorentz resonances involve a
relationship between a particle's orbital frequency and the rate at
which Jupiter (and its magnetic field) rotate. For instance, at the
3:2 Lorentz resonance, a particle completes three orbits for every two
rotations of Jupiter.  Lorentz resonances can have large effects on
small, charged particles. Ring particles can be charged by interacting
with electrons and ions trapped in Jupiter's magnetosphere, with
electrons in Jupiter's extended ionosphere, and via the photoelectric
effect due to sunlight.

The lifetimes of individual particles in the jovian rings are much
shorter than the age of the Solar System, so the particles must be
resupplied by impacts into, and perhaps between, larger parent
bodies. Sputtering, i.e., erosion through impacts of fast ions or
atoms onto solid bodies, is thought to be the most important loss
mechanism for dust particles in the rings. A ``fast drag'' model
\citep{horanyi1996,juhasz2010} predicts much more rapid loss of
individual ring particles, and a somewhat different spatial
distribution of ring material than the work described
above. Observations by the Juno spacecraft, which went into orbit
  around Jupiter in July 2016, may be able to distinguish between the
competing models.

\subsection{Rings of Uranus and Neptune}
\label{subsec:UN}

The rings of Uranus and Neptune, although much less explored than
those of Saturn, have origins that seem easier to understand than
Saturn's rings.  The uranian and neptunian rings are much less massive
than Saturn's and show a complex structure in which numerous small
satellites orbit near the rings. Each ring system contains one or more
narrow, relatively dense ringlets and wide tenuous dust belts. Ringlet
optical depths range from $10^{-6}$ to $\mathcal{O}(1)$ for Uranus and
from $10^{-4}$ to $0.1$ for Neptune. Uranus has 13 small satellites in
the vicinity of its ring system; all except Cordelia orbit outside of
the most opaque ringlet, the $\epsilon$ ring. Cordelia and Ophelia are
thought to shepherd the $\epsilon$ ring
\citep{porco1987,goldreich1987,chiang2000,mosqueira2002}.  Neptune has
6 small satellites near its ring system; four of the six orbit
interior to its densest ring, the Adams ring. The three arcs within
the Adams ring appear to have a resonant relationship with the moon
Galatea \citep{porco1991,sicardy1999,dumas1999,namouni2002}, but the
arcs changed in a decade for reasons that are not well
understood \citep{depater2005}.

The close dynamical association of Uranus and Neptune's rings with the
population of small moons fueled the idea that these rings may derive
from the satellites themselves. In addition, the mass of the uranian
rings is estimated to be comparable to a 10--20~km moonlet for a
density $\simeq 1$~g/cm$^3$, and the neptunian rings are probably much
less massive yet \citep{Esposito_1991}. The ``ring moons'' of Uranus
and Neptune have radii of about 10--100~km. Thus, in principle, there
is enough material in the satellites for them to be a source for ring
material through a process of surface erosion or destruction via
meteoroid bombardment. Incomplete re-accretion due to tides
\citep{Canup_Esposito_1995} may provide a natural explanation for the
coexistence of rings and moons \citep[see also][]{Crida-2015,
  Hyodo_2015}.

In a series of papers, Colwell and Esposito have explored the effect
of meteoroid bombardment on the inner moons of Uranus and Neptune
\citep{Colwell_Esposito_1990b,Colwell_Esposito_1990a,Colwell_Esposito_1992,Colwell_Esposito_1993}. They
show that meteoroid impacts on moons orbiting a giant planet may have
a completely different outcome than, for example, asteroid
collisions. First, the impact velocity of a passing body coming from
orbit around the Sun is very high, typically 20~km/s, due to the
gravitational focusing induced by the giant planet. By comparison, the
average impact velocity in the asteroid belt is about 5 km/s. So every
impact on a moon is much more violent, as the impact energy scales
with the square of the impact velocity. Second, as the small moons
orbit close to their planets, the sizes of their Hill spheres (i.e.,
their spheres of gravitational influence) are comparable to their
physical sizes (see section~\ref{subsec:MARL}). Inside the Roche
limit, the Hill sphere is comparable to or smaller than the physical
size of the body. Just above the Roche limit, the Hill sphere of a
satellite is only a little bigger than the size of the satellite. This
implies that most of the material launched by meteoroid impact is
easily lost because of the tidal field of the host planet, especially
below the Roche limit. As a result, debris produced after the erosion
or destruction of a moonlet will reaccrete with difficulty, or after a
long timescale (many orbital periods). Between destruction of a moon
and re-accretion, debris in orbit around the host planet will slowly
spread longitudinally and will progressively form a ring.

After an impact, the ejecta from a moon with semi-major axis $a$ will
scatter around their ejection point with a radial width $\Delta a/a
\approx V_e/V_{\rm orb}$, where $V_e$ is the ejection velocity,
$V_{\rm orb}$ is the moon's orbital velocity, and $\Delta a$ is the
radial spread of the ejecta. \citet{Colwell_Esposito_1993} compute
that impact ejecta larger than $\approx 1$~cm should gather in
ringlets about 50 km wide, comparable to the observed widths of the
narrow rings of Uranus and Neptune. However, over timescales of many
orbital periods, collisions between fragments will spread the ring. In
addition, effects including Poynting-Robertson drag
\citep{Burns_et_al_2001} and exospheric drag
\citep{Colwell_Esposito_1993,Esposito_1991} may cause the orbits of
ring material to evolve.

\citet{Colwell_Esposito_1992}, \citet{colwell2000}, and
\citet{Zahnle_2003} have estimated the rate of catastrophic disruption
of the moons of the giant planets, assuming the impactors are comets
(``Centaurs''). The model rates are derived by combining crater
size-frequency distributions measured on moons of the giant planets by
Voyager with theoretical extrapolations of the number of comets
observed in the inner Solar System. Under two different assumptions
for the impactors' size distribution, \citet{Zahnle_2003} find that
all of Uranus's inner moons within Puck's orbit should have been
destroyed, on average, in the past $\approx$~0.3 to 2.5 Gyr. The same
calculation for Neptune's satellites shows that all inner moons inside
Proteus's orbit may have been destroyed in the past $\approx$~0.3 to
3.5 Gyr \citep{Zahnle_2003}. These destruction timescales are
estimated by assuming {\em current} impact rates, with a modest
increase as one goes further into the past. The impactor flux may have
been much larger long ago, due either to post-accretional bombardment
just after the formation of the giant planets, or much later, due to
major dynamical instabilities in the Solar System, perhaps some $\sim
650$~Myr after the Solar System formed during the ``Late Heavy
Bombardment'' \citep{Gomes_et_al_2005,charnoz_lhb_2009}.

So Uranus and Neptune's current ring-satellite systems may be in a
cycle of accretion and destruction. Even if the structures we see
today are young (with ages comparable to destruction timescales), the
material they are made of may be as old as the planets themselves. In
addition, because the synchronous orbits of Uranus and Neptune are,
except for Uranus's tenuous $\mu$ ring, beyond their ring systems
(about 3.3 and 3.4 planetary radii for Uranus and Neptune,
respectively), the planet-induced tidal evolution of their satellites
close to the rings is inward. Thus Uranus and Neptune's inner
satellites cannot escape the ring region (which is opposite to the
case of Saturn, for which all satellites finally escape the ring
region, see section~\ref{subsec:MIR}), so this ring-satellite cycle
can continue for as long as there is material in the ring region. For
the uranian rings, the drag from the planet's extended atmosphere
\citep{Esposito_1991} may be the most important mechanism for removing
material from the rings.

However, the current ring systems may, in terms of their total mass,
be insignificant compared to early ring systems that might have
existed around the ice giants. An intriguing recent proposal
\citep{Crida_Charnoz_2012} suggests that Uranus and Neptune originally
had massive rings comparable to Saturn's, and that these rings gave
birth to the regular satellites of the ice giants. This claim is
supported by the similar mass-distance relationship in these three
satellite systems (where the mass grows roughly as the distance to the
Roche limit squared), which fits well with the theoretical
distribution expected from the spreading of rings beyond the Roche
limit.  If true, the rings of Saturn, Uranus and Neptune could have a
similar origin\,; the question then becomes\,: why are they well below
the asymptotic mass described in section~\ref{subsec:Spreading} around
the ice giants and not around Saturn?

The uranian ring-satellite system poses a specific challenge because
of the high obliquity of Uranus ($\approx 98^\circ$; i.e., Uranus is
roughly ``on its side,'' with its rotation poles near its orbital
plane).  In our current understanding of planet formation, it is
unlikely that Uranus formed with such a large obliquity. Uranus's high
obliquity may be due to an event occurring after the planet
formed. Because Uranus's system of rings and satellites lies in its
equatorial plane, either (i) they must have formed after Uranus
acquired its obliquity or (ii) the process that tilted Uranus must
have also been able to tilt its ring and satellite system if they were
already present. Models for tilting Uranus include a collisionless
scenario involving a massive inclined satellite
\citep{Boue_Laskar_2010}, or oblique collisions with nearby
protoplanets \citep[e.g.][]{Morbidelli_et_al_2012}.

In the collisionless scenario, Uranus's spin is slowly tilted by a
resonance between the precession rates of its spin axis and of its
orbital plane, and the satellites adiabatically follow the equatorial
plane of the planet. However, such a resonance requires the presence a
massive satellite ($10^{-3}$--$10^{-2}$ times the mass of Uranus)
  at about 0.01~AU (about $60 R_U$, where $R_U$ is
  the radius of Uranus), compared with $23 R_U$ for Oberon, the
  outermost regular moon) to speed up the known moons'
precession rates \citep{Boue_Laskar_2010}. But this distant satellite
would stay in the ecliptic plane, and would keep the outermost of the
other satellites (Oberon and Titania) in orbits near this plane as
well. This massive satellite, not being observed today, must be
ejected at some point. When this happens, Oberon and Titania would
retain high inclinations relative to Uranus' equatorial plane,
inconsistent with their current extremely low inclinations.

In the collisional scenario, the impact tilts the planet suddenly, but
not the satellite system. The impact also produces a compact, massive
disk of ejecta in the new equatorial plane of the planet
(\citet{Slattery_1992}; called the C-ring by
\citet{Morbidelli_et_al_2012}). This dramatically increases Uranus's
effective $J_2$, and forces a pre-existing proto-satellite disk or satellite
system to precess incoherently about the new equatorial plane. The
randomization of the nodes leads to collisions and damping, so that a
debris disk forms in the new equatorial plane. Note that if the
planet's tilt is more than $90^\circ$, the new debris disk would be
retrograde, so that Uranus must begin with a modest non-zero obliquity,
and thus at least two obliquity-producing collisions are then needed
for Uranus's prograde system. The present satellites and rings of
Uranus then form from this new debris disk, while the satellites
formed by the spreading of the C-ring beyond the Roche limit would be
pulled back inwards by Uranian tides (as they would lie inside the
synchronous orbit) \citep{Morbidelli_et_al_2012}. However, one could
envision that resonant interactions between these satellites born from
the C-ring could push them beyond the synchronous orbit \citep[see,
  for instance,][]{Salmon_Canup_2015DPS}, in which case no
pre-existing proto-satellite disk or satellite system would be needed,
following \citet{Crida_Charnoz_2012}, so long as the tidal $Q_p$
parameter for Uranus was low enough to drive satellites outward to
Oberon's distance.  Then, perhaps Uranus's present rings could be
small remnants of the C-ring, and only one giant impact might be
enough\footnote{Note that the same scenario could apply to Neptune's
  system and its $30^\circ$ inclination.}. In any case, the
disappearance of the C-ring should be studied to understand the origin
of present rings of Uranus.  It is also possible that Uranus once had
massive rings that were entirely lost, with the current rings produced
subsequently.
	
Whereas we seem to understand the close relation between the rings and
satellites, and have identified a cycle of material between the two,
we must ask which came first, the rings or the satellites? This
``chicken and egg'' problem points to {\em the} key question for every
ring formation model: how to put material within a planet's Roche
limit\,? A handful of scenarios have been considered, at least for
Saturn (see section~\ref{subsec:origSat}). For Uranus and Neptune,
only a couple of studies have considered this question. First, a
collision with a nearby protoplanet could do the job
\citep{Slattery_1992,Morbidelli_et_al_2012}.  Another alternative is
through a cometary bombardment during the LHB, but it appears that the
mass injected into the Uranus and Neptune systems may be much higher
than is currently observed \citep{charnoz_chapter_2009}. Thus, for the
moment, there is still great uncertainty on the origin and evolution
of the ring systems of Uranus or Neptune, notably including the
potential for earlier more massive rings.

\section{Other Ring Systems}
\label{sec:other}

\subsection{Centaurs}

The asteroid (10199) Chariklo is the largest known Centaur. Its figure
can be fit with a slightly oblate spheroid of equivalent radius
$127$~km \citep{Chariklo}. Chariklo orbits between Saturn and Uranus,
with a semi-major axis of 15.8~AU and an eccentricity of
0.17. \citet{Chariklo} observed a stellar occultation by this object
from ten observatories in South America.  Before and after the
occultation by the main body of Chariklo, one or two brief extinctions
of the background star were observed. At La Silla, the light curve
had enough time resolution (10~Hz) to clearly see two brief
consecutive occultations on either side of the main body. These
results imply that Chariklo has two rings, with widths of $\approx 7$
and 3~km and optical depths of $\approx 0.4$ and 0.06, separated by
14~km in distance from Chariklo. The rings are about 400~km from the
center of Chariklo. The rings lie near Chariklo's Roche limit if
Chariklo has a density near 1~g/cm$^3$ (the mass of Chariklo is not
well constrained). We refer the reader to Chapter~\ref{Sicardy} for a detailed description of rings around
  Centaurs. 

The existence of such rings is surprising. Although \citet{Chariklo}
show that Chariklo is unlikely to have undergone an encounter so close
to a giant planet that it would have disrupted pre-existing rings, the
question of their formation and evolution remains open. The rings'
apparent confinement suggests that they may be shepherded by moonlets,
as Saturn's narrow F ring is flanked by Prometheus and
Pandora. Unfortunately, our incomplete knowledge of the Chariklo
system, which might have moons or other rings, provides no clue as to
a possible progenitor for the rings, nor for their age. The rings
appear to have sharp edges; unconfined rings would spread rapidly, but
shepherded satellites could maintain sharp edges for a long
time. Narrow rings are also found around Saturn, Uranus, and Neptune,
but it seems difficult to find a formation mechanism that would apply
to giant planets as well as to Centaurs.

\citet{pan_wu_2016} consider three mechanisms for the origin of
Chariklo's rings -- a cratering event that lofted material into orbit
within the Roche limit when Chariklo was still in the Kuiper Belt; an
encounter with a giant planet that perturbed a small moon inward of
the Roche limit; and lofting of dust particles from Chariklo due to
cometary activity. The third mechanism is novel. In this scenario, CO
or N$_2$ outgasses from Chariklo's interior as the body warms during
its journey from the Kuiper Belt to its current orbit between 13 and
19~AU. Seasonal outgassing might lift fine dust from Chariklo's
surface, with a small fraction ending up in stable orbits around the
Centaur. Subsequent collisional evolution of dust and condensates is
required to form the few-meter-sized particles that
\citet{pan_wu_2016} estimate to be present in the rings.  An
interesting aspect of the outgassing model is that it predicts that
rings may be common around Centaurs, whereas they should be uncommon
or absent for KBOs, which are not known to display cometary
activity.  Such an observational prediction may be checked with future
systematic transit observations of KBOs and Centaurs. However, how
outgassing or cratering would produce narrow circular rings is not
clear at all.

\citet{ortiz2015} and \citet{ruprecht2015} reported possible ring
material around another large Centaur, Chiron. Like Chariklo's rings,
the features observed near Chiron were detected by means of stellar
occultations. However, the case that Chiron has rings is not as
certain, because, unlike Chariklo, Chiron displays cometary activity,
so that dips seen away from the Centaur might instead be due to
material ejected from the nucleus, and not orbiting it. A ring system
around yet another large Centaur, Bienor, has recently been suggested,
based on photometry \citep{Fernandez_Valenzuela2016}.

An alternative process was recently proposed to explain the presence
of rings around Centaurs. Using SPH simulations,
\citet{Hyodo_2016_chariklo} showed that a Centaur experiencing a close
encounter with a giant planet (within 1.8 planetary radii) may be
substantially tidally disrupted, but without being fully destroyed. A
fraction of the resulting cloud of debris remains gravitationally
bound to the Centaur as the latter flies away from the giant
planet. The debris then flattens into a ring. This process has the
advantage of being generic. As Centaurs are believed to be objects
from the Kuiper Belt scattered by giant planets inside Neptune's
orbit, such events may have happened in the past.  However, very close
encounters are needed and the rate of such very close encounters is
not yet clear, so the probability of such a scenario still needs to be
evaluated in the future. In addition, the evolution of a debris disk
into a couple of very narrow rings is still unclear. Perhaps satellite
formation occurring inside the debris disk may lead to radial
confinement of material.

\subsection{Exo-rings}

Since the discovery of the exoplanet 51 Pegasi b in 1995, there have
been tremendous efforts to discover new planets, either from the
ground (e.g. HARPS, WASP) or from space (e.g. CoRoT, Kepler,
CHEOPS). Thus far, about 2000 exoplanets have been confirmed, and
thousands of others await confirmation in various observing
programs. Of course, there is, a priori, no reason to believe that
rings are specific to giant planets in our Solar System. We would
naturally expect that rings may be found around exoplanets. However,
the known exoplanets are, in general, quite different from the giant
planets in our Solar System, primarily because detection techniques
still do not allow efficient detection of Solar System-like
planets. These biases may imply that rings are rare around the {\em
  known} exoplanets. Specifically, most planets found via the most
common techniques, radial velocity measurements or transits, are very
close to their central stars, well within 0.5 AU. In addition,
statistically, Earth-sized and Neptune-sized planets are more abundant
than giant planets \citep[see,
  e.g.,][]{lissauer2014,martin2015}. These configurations impose
unusual dynamical and physical conditions on the existence of rings
around confirmed exoplanets.

First of all close to the star we are well inside the snow line, so
that only rings made of refractory material (like silicate minerals)
can survive. Such rings may be very different from Saturn's icy
rings. ``Refractory'' rings will be comprised of denser particles, and
should orbit closer to their host planets, compared to our Solar
System, due to the reduced size of the Roche limit (see
Eq. \ref{eq:Roche}\,) for silicate material.

Second, rings lie in the Laplace plane, that is, the mean orbital
plane above and below which inclined orbital planes precess. This
opens the possibility for warped rings, as the Laplace plane coincides
with the planet's equatorial plane close to the planet, but shifts to
the planet's orbital plane further out. \citet{burns1986} and
\citet{Tremaine_2009} show that the distance below which rings should
lie in the planet's equatorial plane is:
\begin{equation}
R_l=R_p(J_2/q)^{1/5}\ ,
\end{equation}
with $R_p$ standing for the planet's radius, and
$q=(M_\star/M_p)(M_\star/a_p)^3$, with $M_\star$, $M_p$, and $a_p$
standing for the star's mass, planet's mass, and distance to the star,
respectively. Close to the star, a planet may be synchronously
rotating, in which case there is a simple relation between its $J_2$
and $q$: $J_2=\frac{5}{6}k_2q$ \citep{Correia_Rodriguez_2013}. Then,
$R_l$ simplifies to $R_l=R_p(\frac{5}{6}k_2)^{1/5}$. Noting that $k_2$
(the Love number) is, in general, close to 0.5 \citep[see,
  e.g.][]{Yoder_1995}, we find $R_l \approx 0.84R_p$. The conclusion is that
planets orbiting synchronously with their host star would not have
their rings in their equatorial planes, but rather in their orbital
planes\footnote{These two planes may, however, be the same, as
  synchronous rotation is often accompanied by alignment of the spin
  axis with the orbital axis.}. This may make ring detection
by transit difficult, as such rings would be seen edge-on.

Claimed detections of exo-rings are few and are all a matter of debate
because of their intrinsic difficulty. \citet{Kenworthy_Mamajek_2015}
report the detection via transit of an exo-ring system around an
unseen companion of a pre-main sequence K5 star. However, the detected
ring system seems to fill the companion's Hill sphere. Whereas the
structure may indeed exist, its extent and the star's age instead
suggest either a circumstellar or a circumplanetary disk rather than a
typical ring system dominated by tides. The existence of a ring system
around Fomalhaut b (a planet orbiting at about 115 AU from its host
star) has been suggested to explain the unexpected brightness of the
object \citep{Kalas_et_al_2008}. However, to fully account for the
observed brightness, a ring system as wide as 35 planet radii seems
necessary. So, again, this may be rather a circumplanetary structure
than a ring system.

\begin{figure}
\figurebox{}{20pc}{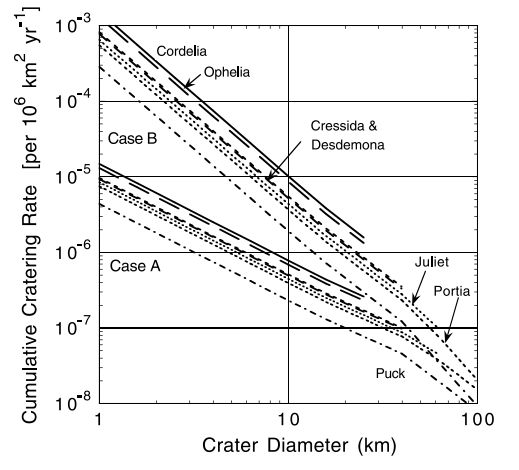}
\caption{Actual bombardment rates of Uranus' inner moons. Adapted from
  \citet{Zahnle_2003}.}
\end{figure}

\begin{figure}
\figurebox{}{20pc}{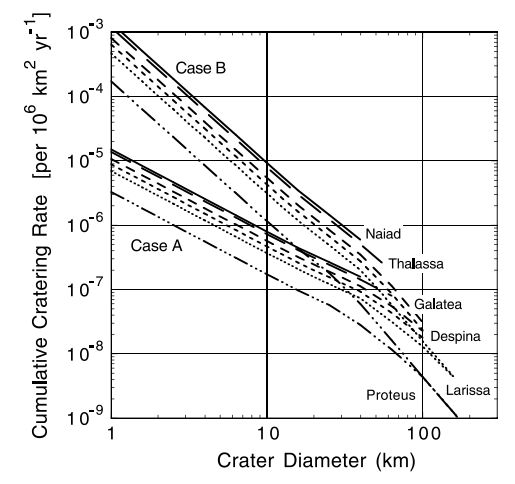}
\caption{Actual bombardment rates of Neptune' inner moons. Adapted
  from \citet{Zahnle_2003}.}
\end{figure}

\section{Conclusion}
\label{sec:conclu}

Do we know how planetary rings are formed? Well, we have to admit that
this is still an open question.  The last three decades of exploration
of our Solar System's planets have shown how diverse and how complex
the different planetary ring systems are. In the past decade, the
close association of ring evolution with satellite formation was
identified and allowed us to make significant progress in our
understanding of rings. Whereas we are still not sure of the process
of ring formation, we envision several possible different scenarios,
but each faces its own problems. The most difficult questions to
address include\,:

\begin{itemize}
\item Why are the rings around the four giant planets so different?
\item Why do they all revolve in the prograde direction?
\item Why are Saturn's rings so ice-rich, while those of Uranus and
  Neptune seem to be made of a non-icy material?
\item Why are Saturn's rings orders of magnitude more massive than the
  other ring systems, despite potentially common formation processes?
\item Why don't terrestrial planets have rings?
\end{itemize}

The challenge is creating a ring formation scenario that allows enough
variability to explain the wide diversity of planetary rings. Perhaps
there is no unique scenario to form rings, so that every planet may
have a different history. For the case of Saturn, it seems that tidal
stripping from the outer ice shell of an ancient differentiated
satellite, collisional disruption of a pre-existing small moon, or
tidal disruption of the outer icy layers of a differentiated planetary
interloper may all represent viable formation mechanisms in case the
rings are as old as Saturn itself. But when and how such events
occurred is still uncertain. For tidal stripping from an inwardly
migrating satellite to produce a long-lived ring, there must be a
favorable agreement in timing between satellite migration and the
disappearance of the circumplanetary disk in order for the ring to
survive against gas drag. Perhaps Saturn's ring was the only one to
achieve such conditions, with thus-formed massive rings at the other
planets destroyed. If Saturn's rings formed by the total disruption of
a small moon, early tidal evolution at Saturn must have been slow, and
an explanation is needed as to why the rings (and many of the
medium-sized inner satellites) are so ice-rich, rather than a mixture
of rock and ice as expected for solar abundances. \ac{We still await
  the results of the CDA instrument on Cassini, which will help to
  constrain the pollution timescale of Saturn's rings.} If Saturn's
rings formed by the tidal disruption of the outer layers of
differentiated comets or large KBOs that made close passages during
the LHB, then we see no reason why the outer planet ring systems would
all orbit in the prograde direction, nor why the other giant planets
would lack massive ring systems today.

Whereas the exact timing of these scenarios is unknown, they all imply
that either the rings formed concurrently with Saturn or that they
formed during a phase of intense bombardment of the outer Solar
System, the so-called Late Heavy Bombardment that may have happened
about 3.8 Gyr ago. Forming rings in a much more recent period seems a
difficult task, as we do not know any major restructuring event of the
Solar System in, at least, the last 3~Gyr. However, \citet{cuk2016}
emphasize that evection resonances of some of Saturn's moons with the
Sun may \ac{have lead recently to} resonantly-driven collisions in the
system. During \ac{such destructive} collisions among satellites,
massive disks of debris may form and may re-accrete into a new
generation of moons and, putatively, could \ac{create a young} ring
system. This scenario has not been tested numerically and we view it
as unlikely that it will be able to reproduce the observed
mass-distance distribution of Saturn's moons as well as the ice-rich
composition of the rings.

Perhaps for Uranus and Neptune giant impacts played an additional role
in modifying the final systems. We think that the Earth suffered a
giant impact that gave birth to our Moon via a Roche interior disk
\ac{\citep{Cameron-Ward-1976,Kokubo-etal-2000,Canup-Asphaug-2001,Crida_Charnoz_2012,Charnoz-Michaut-2015}},
and perhaps Mars' small moons formed in a similar way
\citep{Rosenblatt_2012,Rosenblatt_2016}. Terrestrial rings may have
been lost due to the stronger effects of the Sun's radiation at their
smaller semi-major axes, but this has yet to be explored. \ac{The
  perturbations from the Earth's very massive Moon
  \citep{Cameron-Ward-1976,Kokubo-etal-2000}, and the inwards tidal
  migration of Mars's satellites \citep{Rosenblatt_2016} could also
  explain how these two planets lost completely their rings.}

Developments since 2010 have shown that the rings' evolution, since it
is linked to the satellites' evolution, is also coupled to the
planet's internal structure, as it is the latter that controls the
intensity of tidal dissipation. So it seems that the rings,
satellites, and planet form a single system with a strong degree of
coupling between its different components.  Juno, now in orbit at
Jupiter to study the planet's interior, or JUICE, which is scheduled
to go into orbit around Jupiter in 2030, will undoubtedly provide
invaluable information on the Jupiter system, which we hope will, in
turn, place new constraints on the origin of planetary rings.

Finally, whereas exo-rings have still not been clearly detected, they
should be discovered in the next few years. Exo-rings may be detected
in transit if the host planet does not orbit too close to its
star. The CHEOPS and PLATO missions, with launches planned for 2018
and 2024, may detect such structures. If exo-rings are found around
some type of exoplanets, and not others, this may again put
constraints on their origin. In the near future, the James Webb Space
Telescope, which should be launched in 2018, will have the capability
to image rings around exoplanets far from their parent
stars. Undoubtedly, our understanding of the origin of planetary rings
should advance substantially in the next few years.

\section{Acknowledgements}

SC acknowledges the financial support of the UnivEarthS Labex program at Sorbonne Paris Cit\'e (ANR-10-LABX-0023 and ANR-11-IDEX-0005-02). RMC acknowledges support from NASA's Planetary Geology and Geophysics program. LD thanks the Cassini project and NASA’s Outer Planets Research program for support.

\bibliographystyle{cambridgeauthordate}
\bibliography{origin_chapter_robin_luke_DEC_2016}\label{refs}

\vfill\eject

\newcounter{chapterdummy}
\newenvironment{chapterdummy}{\refstepcounter{chapterdummy}}

\textcolor{white}{
\chapterdummy{Space age studies of planetary rings \label{Esposito}}
\chapterdummy{An introduction to planetary ring dynamics \label{HedmanIntro}}
\chapterdummy{The rings of Saturn \label{Saturn}}
\chapterdummy{The rings of Uranus \label{Uranus}}
\chapterdummy{The rings of Neptune \label{Neptune}}
\chapterdummy{The rings of Jupiter \label{Jupiter}}
\chapterdummy{Rings beyond the giant planets \label{Sicardy}}
\chapterdummy{Moonlets in dense planetary rings \label{Spahn}}
\chapterdummy{Meteoroid bombardment and ballistic transport in planetary rings \label{Estrada}}
\chapterdummy{Theory of narrow rings and sharp edges \label{Longaretti}}
\chapterdummy{Narrow rings, gaps, and sharp edges \label{NicholsonEdges}}
\chapterdummy{Dusty rings \label{HedmanDusty}}
\chapterdummy{The F ring of Saturn \label{Murray}}
\chapterdummy{Plasma, neutral atmosphere, and energetic radiation environments of planetary rings \label{Cooper}}
\chapterdummy{Thermal properties of rings and ring particles \label{SpilkerThermal}}
\chapterdummy{Computer simulations of planetary rings \label{Salo}}
\chapterdummy{Laboratory studies of planetary ring systems \label{Colwell}}
\chapterdummy{The origin of planetary ring systems \label{Charnoz}}
\chapterdummy{Future missions to planetary rings \label{SpilkerMissions}}
\chapterdummy{Planetary rings and other astrophysical disks \label{Latter}}
\chapterdummy{The future of planetary rings studies \label{Tiscareno}}}

\end{document}